%% file: main.tex
\documentclass{article}
\usepackage[utf8]{inputenc}

\usepackage{tikz}
\usetikzlibrary{matrix,arrows,decorations.pathreplacing,tikzmark,calc}
\tikzset{break/.style={fill=white, inner sep=1pt}}
\tikzset{inline/.style={column sep = 2.5em,inner sep = 1pt}}
\usetikzlibrary{arrows,positioning}

\usepackage{amsmath,amsfonts,amsthm,amssymb,thmtools,mathtools,mathrsfs,dsfont}
\usepackage{bbold}
\usepackage{graphics}
\usepackage{caption}
\usepackage{subcaption}
\usepackage{epstopdf}
\usepackage{enumerate}
\usepackage{enumitem}
\usepackage{etex}
\usepackage{setspace}
\usepackage{chngpage}
\usepackage{soul}
\usepackage{ifthen}
\usepackage{indentfirst}
\usepackage{listings}
\usepackage{url}
\usepackage{longtable}
\usepackage{rotating}
\usepackage{float}
\usepackage{microtype}
\usepackage{lineno}
\usepackage[caption = false]{subfig}
\usepackage[margin=1in]{geometry}
\allowdisplaybreaks

\usepackage[labelfont=bf]{caption}
\usepackage{multirow}

\usepackage{array}
\newcolumntype{L}[1]{>{\raggedright\let\newline\\\arraybackslash\hspace{0pt}}m{#1}}
\newcolumntype{C}[1]{>{\centering\let\newline\\\arraybackslash\hspace{0pt}}m{#1}}
\newcolumntype{R}[1]{>{\raggedleft\let\newline\\\arraybackslash\hspace{0pt}}m{#1}}

\theoremstyle{plain}

\theoremstyle{definition}

\title{Edge-based Modeling for Disease Transmission on Random Graphs: An Application to Mitigate a Syphilis Outbreak}
\author{S. Zhao, S. Saeed, M. Carter, B. Stoner, M. Hoover, H. Guan and F.M.G. Magpantay}

\date{}

\begin{document}

\maketitle
\begin{abstract}
\bigskip
Edge-based network models, especially those based on bond percolation methods, can be used to model disease transmission on complex networks and accommodate social heterogeneity while keeping tractability. 
Here we present an application of an edge-based network model to the spread of syphilis in the Kingston, Frontenac and Lennox \& Addington (KFL\&A) region of Southeastern Ontario, Canada. 
We compared the results of using a network-based susceptible-infectious-recovered (SIR) model to those generated from using a traditional mass action SIR model.
We found that the network model yields very different predictions, including a much lower estimate of the final epidemic size.
We also used the network model to estimate the potential impact of introducing a rapid syphilis point of care test (POCT) and treatment intervention strategy that has recently been implemented by the public health unit to mitigate syphilis transmission.
\end{abstract}

\section{Introduction}
\label{sec: Intro}

\subsection{Syphilis}
\label{sec: Syphilis}
Syphilis is a sexually transmitted and blood-borne infection (STBBI) caused by the bacterium \textit{Treponema pallidum} subspecies pallidum \cite{DeSchryverMeheus:1990, GhanemHook:2020}.
Typically, a one-time intramuscular injection of \textit{benzathine penicillin G.} can effectively treat early syphilis infections (less than one-year duration)~\cite{SinghRomanowski:1999,ValentineBolan:1999}.
Left untreated, syphilis can have very serious health consequences and move through the following four stages:

\begin{itemize}
\item Primary Stage: Symptoms may include swollen glands and one or more painless chancres~\cite{PHACSyphilis}, typically appear around 3 weeks after infection. 
This stage usually lasts 9--90 days~\cite{French:2007, MackrellEtAl:2024} and the chancre heals regardless of treatment~\cite{CDCSyphilis:2023}.

\item Secondary Stage: Symptoms include fever, headache, patchy hair loss, swollen glands in the groin or neck, condylomata lata and rashes.
These symptoms can develop 4-10 weeks after exposure and then symptoms last for 3-12 weeks~\cite{French:2007, MackrellEtAl:2024}.

\item Latent Stage: There are no noticeable symptoms. This can last for up to 20 years.
Cases in this stage can be further separated into two sub-classes based on infectious duration.
\begin{itemize}

\item Early latent stage: No symptoms but infected with syphilis for less than 12 months. 
At this stage the disease can still be transmitted to others \cite{PHACSyphilis}.

\item Late latent stage: No symptoms but infected for more than 12 months. 
Late latent stage syphilis is typically not transmissible to others \cite{PHACSyphilis, CDCSyphilis:2023}.

\end{itemize} 

In practice, diagnoses of early versus late latent stages are determined based on serological testing and clinical input.

\item 
Tertiary Stage: Syphilis invades significant organ systems like the brain, heart and blood vessels.
Untreated tertiary syphilis can eventually lead to death.
\end{itemize}

\begin{figure}[htbp]
    \begin{center}
        \includegraphics[width=0.95\textwidth]{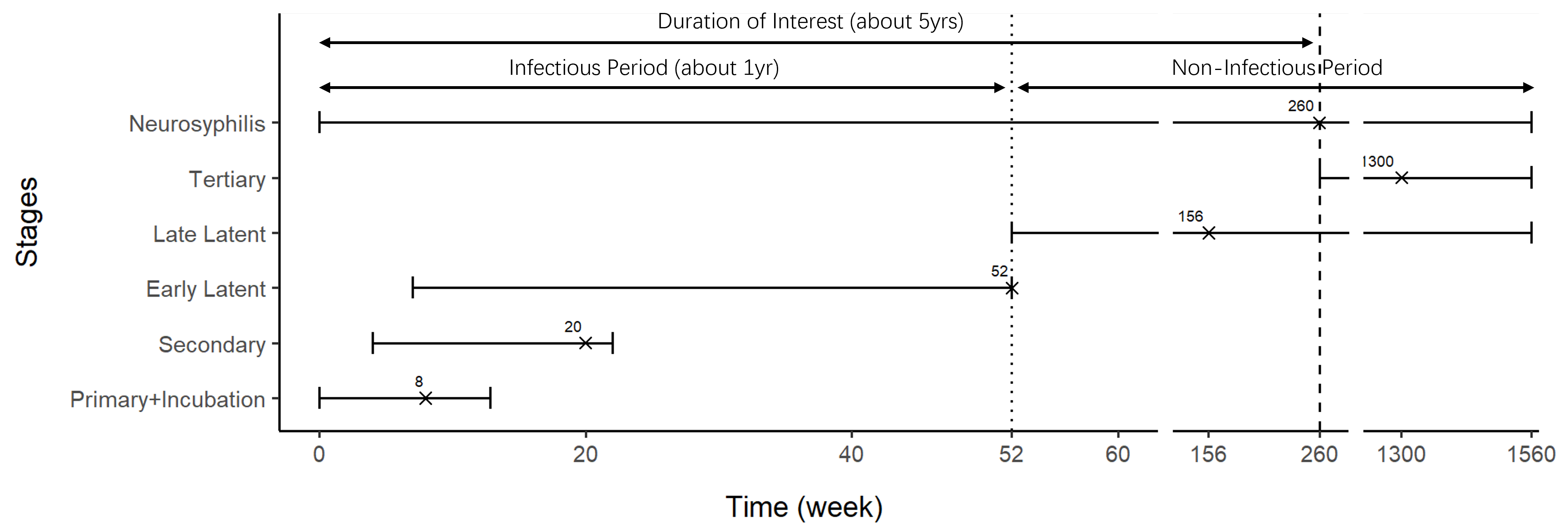}
    \end{center}
    \caption[Time frame of syphilis stages]{Time frame of syphilis stages after infection. 
    The crosses represent the typical observed duration of each stage, determined from consultation with local public health professionals, which we use as the expected duration of each stage in our modeling work.
    For neurosyphilis, the typical value is taken to be at the maximum duration of interest of our project, which is 260 weeks (about 5 years).}
    \label{fig: Time Frame}
\end{figure}

Neurosyphilis occurs when syphilis infection reaches the central nervous system, and can occur at any stage of the disease~\cite{PHACSyphilis,CDCSyphilis:2023}.
The typical duration of each stage is based on the literature and consultation with local public health professionals.
These ranges are important for our modeling work and are summarized in Figure~\ref{fig: Time Frame}.

Congenital syphilis occurs when a pregnant woman with untreated syphilis passes the infection to her baby during pregnancy or at birth. 
This can result in severe adverse pregnancy outcomes including miscarriage, stillbirth and infant death shortly after birth. 
Infected infants can also be born seemingly healthy but later on develop serious problems including nerve damage and hearing loss~\cite{PHOCongenital:2022}. 

Despite record low syphilis rates in the 1990's, the infection has resurfaced as a growing public health issue in Canada, with a 109\% increase in reported cases from 2018 to 2022~\cite{AhoEtal:2022}. 
In Ontario, the largest province by population, the number of syphilis cases between 2013 and 2022 grew from 3.9 to 23.6 cases per 100,000 people~\cite{PHOSyphilis:2024}.
Women aged 15 to 39 were identified as a growing at-risk population \cite{PHOSyphilis:2024, PHOCongenital:2022}, which has also led to congenital syphilis cases increasing, going from 6 cases in 2021 to 26 cases in 2022~\cite{PHOCongenital:2022}.
A significant proportion of syphilis infections have been associated with underserved groups with less access to medical resources and increased risk factors. 

The Kingston, Frontenac, and Lennox \& Addington Public Health  (KFL\&A PH) unit reported a syphilis outbreak in December 2022 when the rates of infectious syphilis (41.2 per 100,000) surpassed the provincial average (23.6 per 100,000)~\cite{PHOSyphilis:2024}. 
Five congenital cases (2.8 cases per 1000 live births) were also reported that year, which is one of the highest rates within the 34 public health units in Ontario.
As part of a strategic response to this outbreak, KFL\&A PH implemented a rapid test and treat protocol using a recently approved point-of-care test (POCT) in Canada.
This public health intervention allowed trained healthcare providers to screen for syphilis infections in conjunction with existing outreach services and immediately treat persons who screened positive for syphilis. The goal was to reduce the time between traditional test and treat with serology, and ultimately reducing the number of people who become lost to follow-up. 
More details on the intervention are available in Mackrell et. al.~\cite{MackrellEtAl:2024}.

The impact of this rapid test and treat intervention will be important to inform public health policy in areas experiencing similar outbreaks.
Here we present a method to model syphilis transmission and estimate the impact of this intervention using edge-based percolation techniques.

\subsection{Percolation Network}
\label{sec: Perco Net}

Mathematical models of infectious disease spread are often based on the classical susceptible-infectious-recovered (SIR) framework originally introduced by Kermack and McKendrick \cite{KermMcKe}. 
Many implementations of the SIR models come with an assumption of homogeneous mixing of individuals in the population yielding a ``mass action'' (MA) term in the resulting system of ordinary differential equations. 
We refer to these traditional SIR models with a mass action assumption as MA-SIR models.
These have been widely used for many types of diseases, since this assumption keeps the dynamics of the model tractable for analysis.
However, the MA assumption can be problematic for modeling sexually transmitted infections where assuming homogeneous mixing can cause overestimates of the actual outbreak size.
There have been many studies that attempt to eliminate the MA assumption without significantly sacrificing analytical tractability.
Newman, Strogatz and Watts~\cite{NewmanStrogatzWatts:2001} developed the basic network percolation model for epidemiology based on random simple graph theory introduced by Bollobas~\cite{Bollobas:1980}.
For these models, vertices in a  network represent individuals in the community, and edges between vertices represent potentially disease-causing connections.
The number of edges connected to each vertex is called the degree of the vertex.
For a sexually transmitted infection, the relevant network would be the sexual contact network and degrees would represent the number of sexual partners.
The social structure is represented by degree sequences, and predictions can be made on disease spreading by computing the expectation among all possible graphs with such degree sequences.
In further work, Newman~\cite{Newman:2002} generalized and optimized the method using the bond percolation model by Frisch and Hammersley~\cite{frisch:1963} and Grassberger~\cite{grassberger1983critical}.
We refer to this generalized method as the typical percolation method.

Percolation theory can be applied to model outbreaks on epidemic networks with generalized degree sequences that reflect more diversity in social structures among different communities.
Moreover, these methods also provide an expression for transmissibility, and allows the model to work with a distribution of key epidemiological parameters.
Even under such generalization, the method manages to retain a relatively simple form for analysis and forecasting.
This method has gained a lot of attention as one of the most significant networks approaches for epidemiology models. 
It has also been modified to fit more complicated assumptions, such as partial immunity~\cite{BansalMeyers:2012}.

The typical bond percolation method has two significant limitations: it lacks dynamical information about the outbreak, and it can only be applied to static configuration networks without network changes.
But its idea of using percolation theory inspired many other approaches for modeling epidemics in networks. 
One of the most well-known is the method first introduced as a pair-approximation model by Keeling~\cite{Keeling:1999}.
This model was then generalized by Volz~\cite{Volz:2008}, and further formalized by Miller, Slim and Volz~\cite{MillerSlimVolz:2012}.
We refer to the Miller, Slim and Volz~\cite{MillerSlimVolz:2012} model as the modified percolation model.
This constructs a connection between the network patterns and a relatively tractable dynamical system, which provides dynamical information about the outbreak.
In a previous paper we proved that the results are equivalent to Newman's method on configuration networks~\cite{ZhaoMagpantay:2024}, and can be applied to some networks with evolutionary changes with reasonable tractability.
More details about the theoretical background, definition, assumptions of the edge-based model and the connection between typical and modified percolation models can be found in the author's previous work~\cite{ZhaoMagpantay:2024}.

To model syphilis transmission we use the modified percolation model with an SIR-type disease progression. 
In this paper we refer to this as the network-SIR model.
We compare the predictions provided by the network-SIR model with those from more traditional MA-SIR models.

\section{Data}
\label{sec: data}
\begin{figure}[htbp]
    \centering
    \begin{subfigure}[b]{0.32\textwidth}
         \centering
         \includegraphics[width=\textwidth]{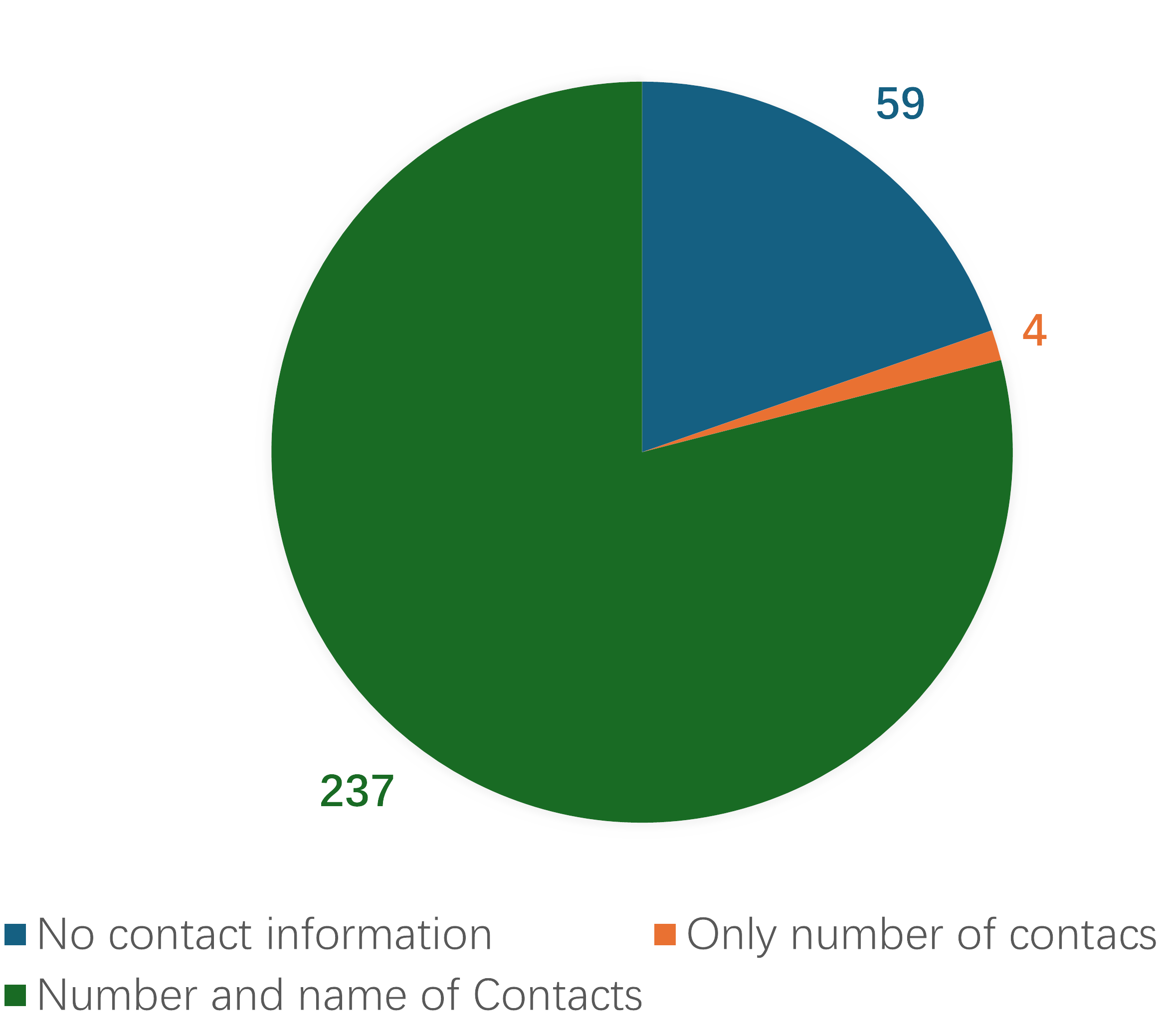}
         \caption{Proportion of contact information}
    \end{subfigure}
    \begin{subfigure}[b]{0.30\textwidth}
         \centering
         \includegraphics[width=\textwidth]{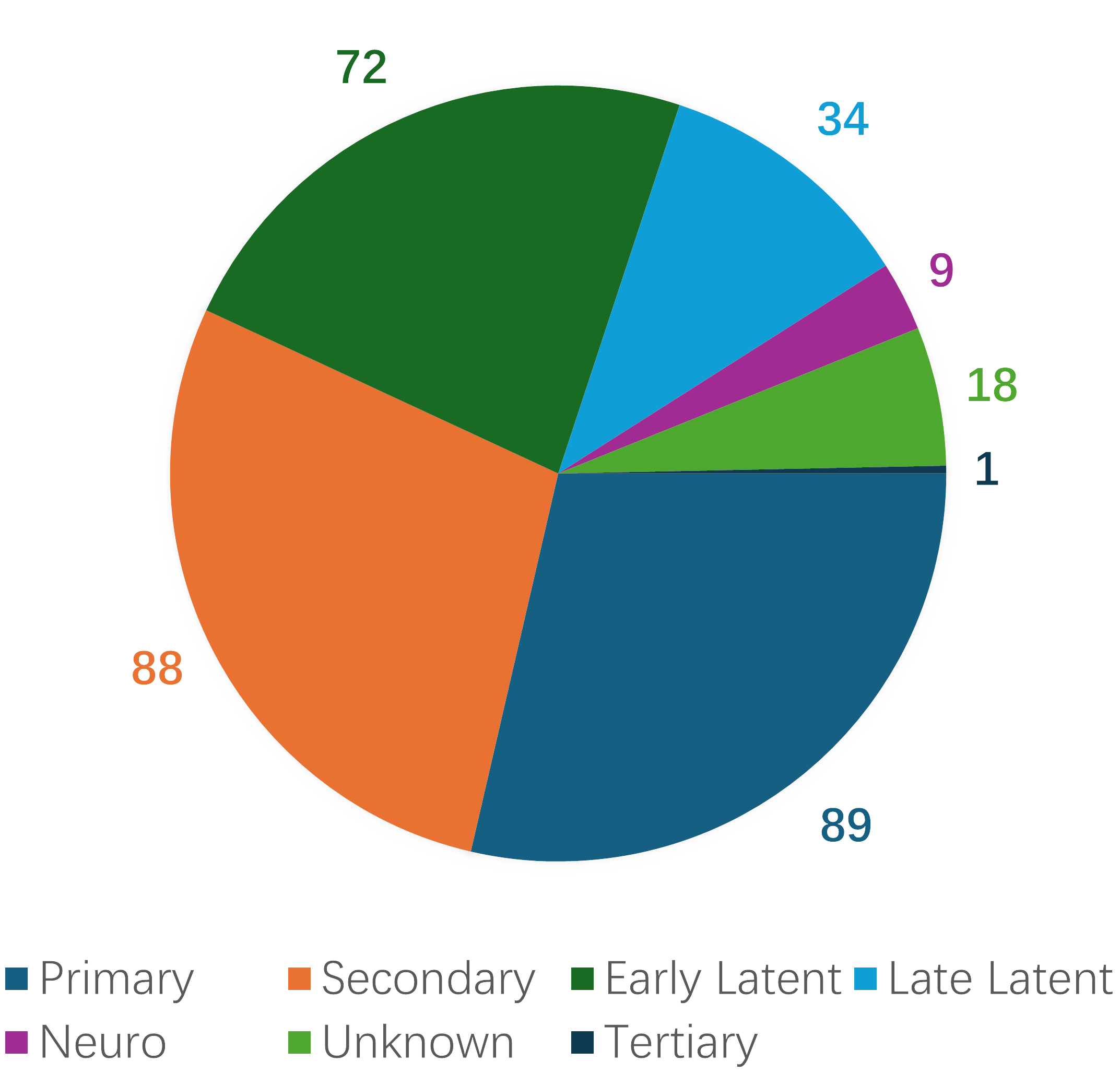}
         \caption{Proportion of syphilis Stages}
    \end{subfigure}
    \begin{subfigure}[b]{0.33\textwidth}
         \centering
         \includegraphics[width=\textwidth]{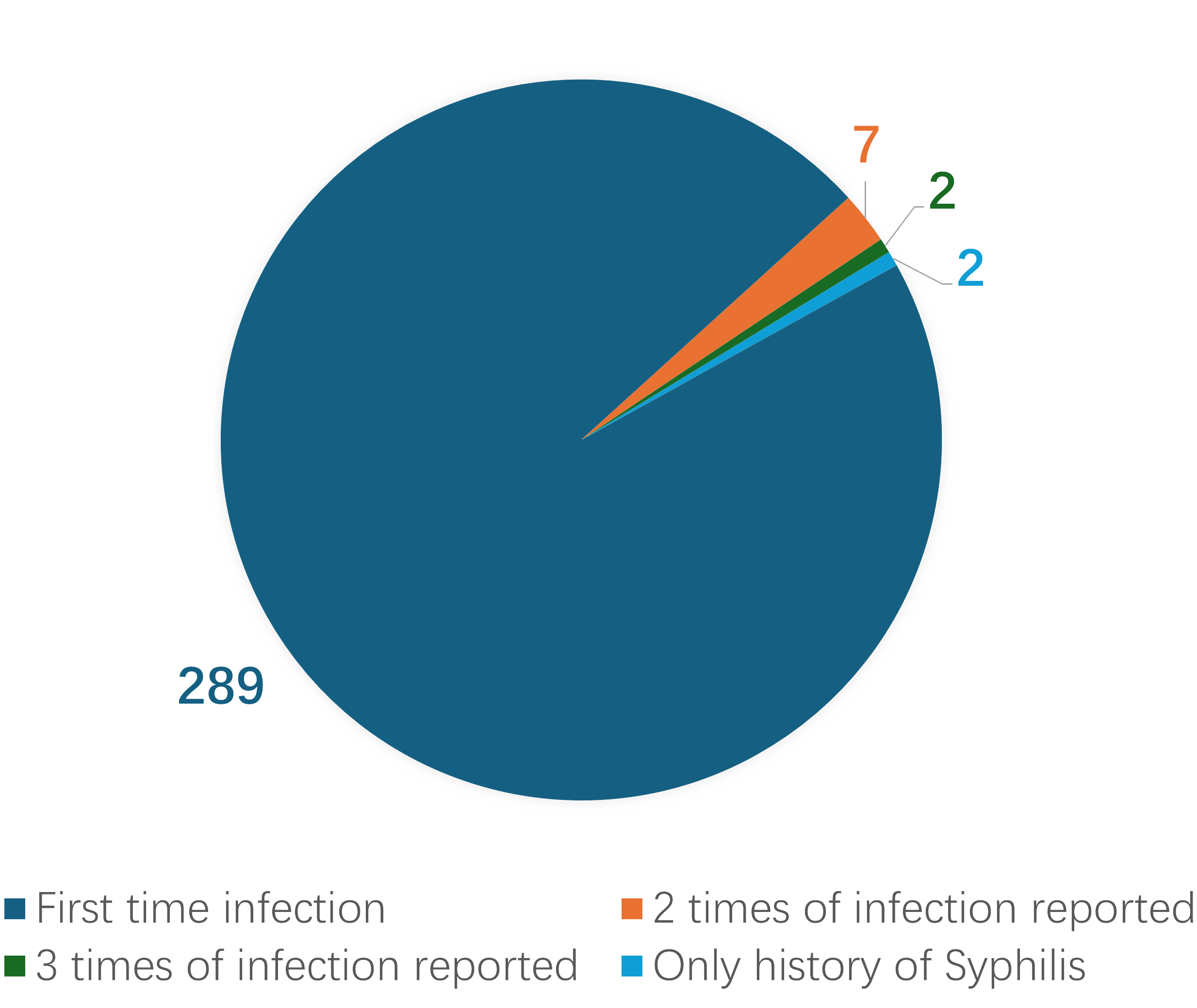}
         \caption{Proportion of Reinfection}
    \end{subfigure}
    \caption{Proportions of adult case data}
    \label{fig: proportions}
\end{figure}

We obtained KFL\&A PH syphilis case data collected from January 2019 to December 2023, which included 306 reported cases, with information on the date of encounter/diagnoses, age at the time of encounter, syphilis stage, medical, social and environmental risk factors, and contacts with potential for transmission in past 6 months (P6M) provided by cases.
The reported cases consisted of 300 adult cases and six newborn cases.
As we can see in Figure~\ref{fig: proportions}(a), among the adults, 237 (79.0\%) of them provided both the number of transmissible contacts in the P6M with names, four (1.3\%) provided only their number of contacts and 59 (19.7\%) did not provide any information on their contacts.
Figure~\ref{fig: proportions}(b) shows the stage distribution of syphilis among reported adult cases.
Moreover, Figure~\ref{fig: proportions}(c) illustrates the proportion that reported reinfection of syphilis, with 289 (96.3\%) of the cases having had syphilis once during the monitoring period, seven (2.3\%) twice, two (0.7\%) three times, and two more (0.7\%) only reporting having it once during the monitoring period but with a history of syphilis before monitoring.

A plot of the time series of the number of reported syphilis cases is shown in Figure~\ref{fig: MSV R Cumulative} together with the trajectories from the fitted models.
Since the KFL\&A PH unit implemented the rapid test and treat protocol in June 2023, 
we only use the first 116 bi-weeks (terminating on 29 May 2023) for model fitting.
Throughout this paper we set time zero to be 1 January 2019 and plot time in units of bi-weeks.

\section{Model Setup}
\label{sec: model}
We constructed a network-SIR model, as described in Section~\ref{sec: Perco Net}, to evaluate syphilis transmission within the underserved population in the KFL\&A area.
This model serves as the base for understanding the transmission dynamics of the disease and its interaction with the community structure before POCT implementation.
The construction of this base model consisted of two phases: the first is to generate representative random networks, characterized by their size and degree distribution, of the community. 
This will be discussed in Sections~\ref{sec: size}--\ref{sec: Net&Deg}.
The second phase is the use of the modified percolation process discussed in Section~\ref{sec: Perco Net} and fitting this to the time series data on syphilis cases. 
These results will be presented in Section~\ref{sec: Epi Dyn}.
The results derived using a network-SIR model were compared against those from a traditional MA-SIR model in Section~\ref{sec: MA compare}.

\subsection{Estimating the Size of the Target Population}
\label{sec: size}
The KFL\&A PH unit identified the underserved communities of people experiencing homelessness or having housing instability, and people using drugs as a high-risk population.
Even though individual-level data on each case's social and economic status is not routinely collected, this correlation is evident from the high incidence of related risk factors.
Therefore, we constructed the network-SIR model based on the collected data from the cases' reported contacts.
We supplemented this data with information from a literature review and consultation with local public health experts.

\begin{figure}[htbp]
    \centering
    \begin{subfigure}[b]{0.30\textwidth}
         \centering
         \includegraphics[width=\textwidth]{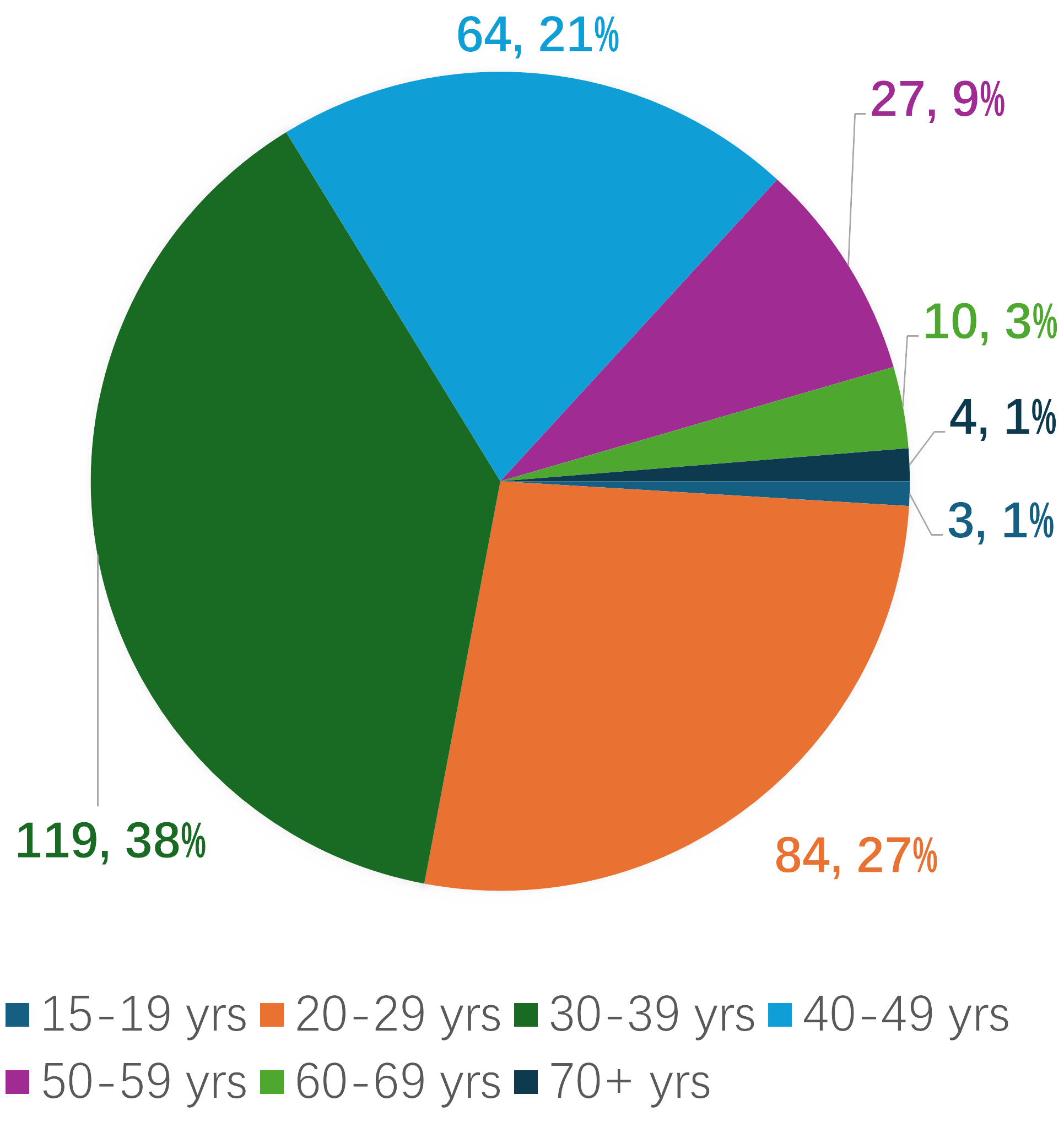}
         \caption{Proportion in KFL\&A 2019-2023 (N=311)}
    \end{subfigure}
    \begin{subfigure}[b]{0.31\textwidth}
         \centering
         \includegraphics[width=\textwidth]{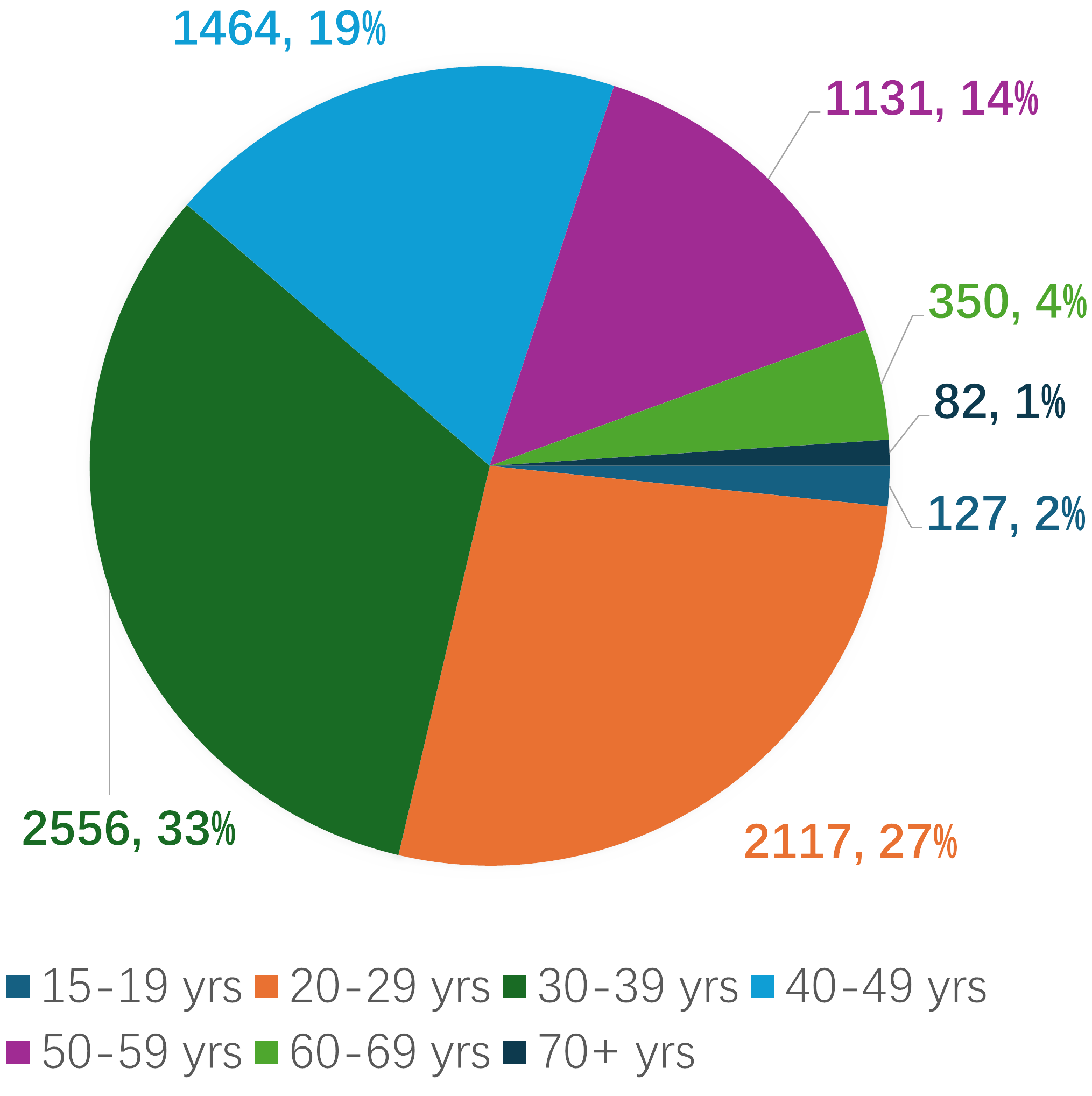}
         \caption{Proportion in Ontario 2019-2021 (N=7827)}
    \end{subfigure}
    \begin{subfigure}[b]{0.33\textwidth}
         \centering
         \includegraphics[width=\textwidth]{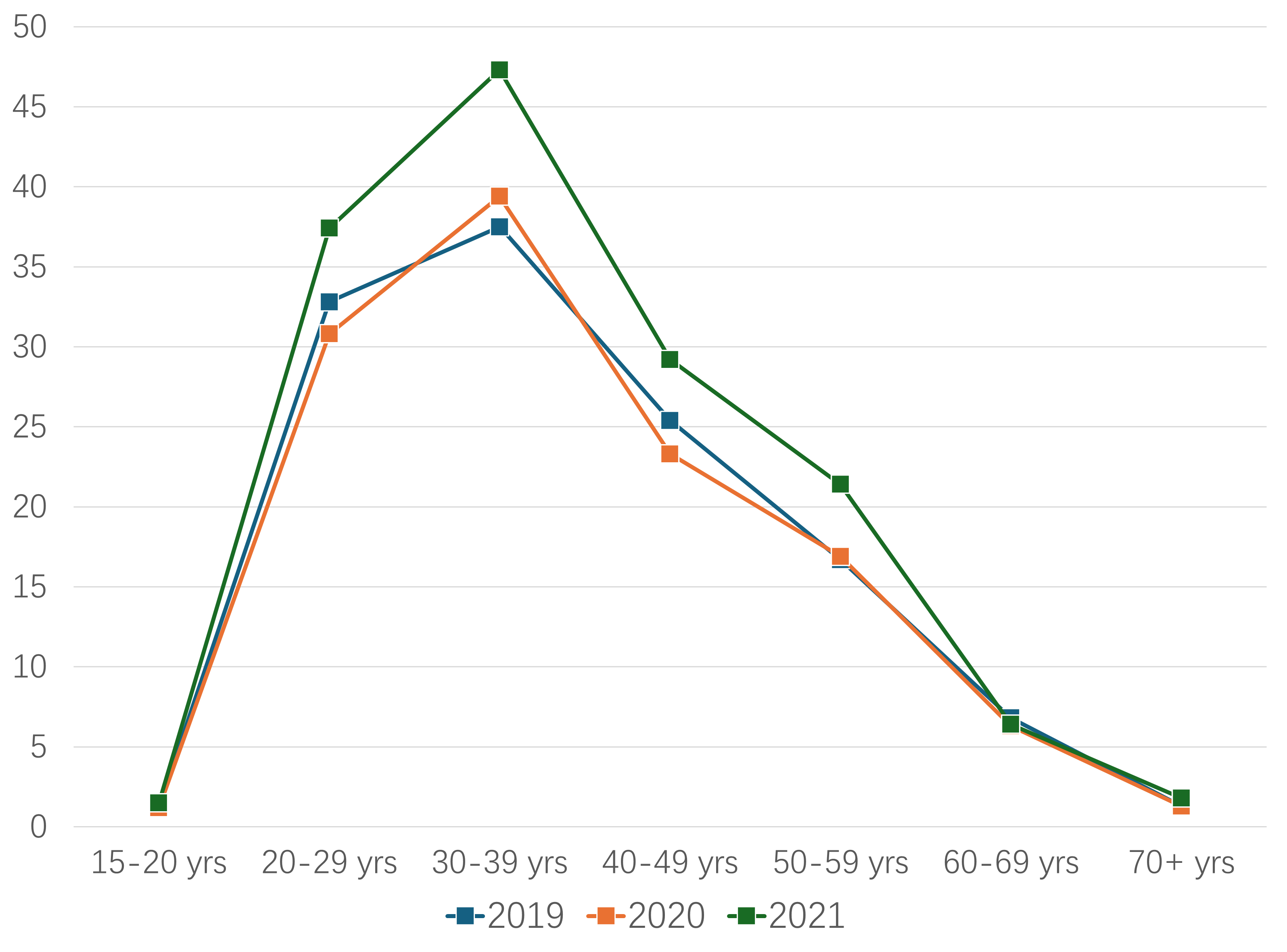}
         \caption{Annual rate per $100,000$ population in Ontario, 2019-2021 }
    \end{subfigure}
    \caption{Case Proportions and Rates by Age}
    \label{fig: Age}
\end{figure}

To estimate the size of the relevant network of individuals at risk of syphilis infection within the KFL\&A area, we made the following considerations:
The size of the population served by KFL\&A PH was 214,513 in 2023.
However, it is clear that only a subset of that population is at an elevated risk of syphilis transmission.
We refer to this subpopulation as the \textit{target population}. From consultation with KFL\&A PH\, we make the assumptions that the individuals in this target population have the following characteristics:
\begin{enumerate}
    \item [(a)] Financially or socially constrained, so they have reduced access to public health services.
    \item [(b)] Have at least one direct or indirect potentially disease-causing connection with other members of the at-risk population.
\end{enumerate}
Since we expect that the majority of the transmission in KFL\&A is via sexual contact, one way to incorporate both (a) and (b) is to estimate the number of individuals in the financially or socially constrained subpopulation that are also sexually active. 
Specifically, to take into account assumption (a), we restrict to the fraction of population with \textit{low-income measure after tax (LIM-AT)} as defined by Statistics Canada~\cite{STATCAN}.
We took assumption (b) into account by restricting to individuals between ages 18 to 64 since this is considered to be an age range that includes most sexually active individuals.
This choice is supported by the local and provincial data~\cite{PHOSyphilis:2024} as shown in Figure~\ref{fig: Age}, which also shows the concentration of syphilis cases in these age ranges.

Based on economic data and demographic data, approximately $10.6\%$ of the population qualify as low income (using LIM-AT) and are between the ages 18 to 64.
KFL\&A PH also noted a small spillover of syphilis transmission to the local medium/high-income community so we assumed that the target population is slightly larger, at $12.0\%$ of the population, yielding approximately $26,000$ individuals.
Thus, we set the size of the network $N=26,000$.

We note that age and economic status were only used to estimate the target population size.
The random network model is originally built on networks with infinite size, and all predictions are given as proportions of population.
Therefore the population size is only required for fitting the model to the data (mainly affecting estimates of the reporting probability), but has limited impact on the dynamical behaviour of the model.


For further simplicity and tractability of the model equations, we need to make several more assumptions about the target population. To make these assumptions explicit we clearly label them as (H1)--(H10). 
The first assumption is given below:
\begin{enumerate}
\item [(H1)] No individual enters or leaves the target population during the time duration of interest.
\end{enumerate}
Assumption (H1) is a strong assumption to impose on the model, but it is necessary for a simple approach to our first pass at fitting a network-SIR model to the data.
Under assumption (H1) there is no need to consider birth, death and immigration. 
The size of the target population is assumed to be constant.
The general difficulty in tracking changes in relationships among the underserved community, due to many issues such as the lack of permanent addresses, avoidance of administrative services, etc., makes (H1) a necessary assumption at this point.
It is also not an unreasonable assumption due to the following reasons:
\begin{itemize}
\item We fit the data over a relatively short time period (2019-2023) so that the change in the population due to births and natural death is relatively small.
\item Syphilis is curable and has a low direct death rate, especially in the age group we are looking at, so the impact of death due to disease is limited.
\end{itemize}

\subsection{Degree Distribution}
\label{sec: Net&Deg}

The next step to build the network-SIR model is to construct a degree distribution of the target population.
To do this, we require a few more assumptions.
\begin{enumerate}
\item [(H2)] For each vertex, its degree is the number of other vertices that could have had transmissible contact with it during the whole duration of interest. 
\end{enumerate}
By making assumption (H2) we assume that the network represents all contacts within the whole duration of interest, instead of just contacts in any short period of time. 
Together, assumptions (H1)--(H2) allows us to construct the network as configuration model (CM).
Such assumptions simplify the model and make it close to the modified percolation model~\cite{MillerSlimVolz:2012} discussed in Section~\ref{sec: Perco Net}.
It also makes the base model more comparable with MA-SIR models, where contact rates are also assumed to be invariant with respect to time.

The degree distribution of the network is assumed to be the distribution of the number of sexual contacts for the target population.
Since edge-based percolation techniques are based on degree distributions instead of any specific network structure, we do not need the actual social structure of the community.
We only need to determine the degree distribution. 
In the absence of other data on this, we decided to treat the reported cases as a random sample of the target population.
While this would clearly be a biased sample of the general population, the bias is reduced if we only consider the target population.
\begin{enumerate}
\item [(H3)] The cases in the data who reported their contacts represent an unbiased random sample from the target population.
\end{enumerate}

The data includes 237 cases with a contact list within P6M of diagnosis. 
It also includes four cases who were only able to provide their estimated numbers of sexual contacts in the P6M.
The rest of the cases did not provide information on their contacts.
The data also contains the risk factors associated with each case, including sexual, behavioural, medical, pregnancy-related, financial and social status.
This list includes factors that are commonly considered to have a significant impact on the risk of syphilis infection by provincial public health\cite{PHOSyphilis:2024} and additional information KFL\&A PH determined important given the local context.
A complete list of risk factors is shown in Table~\ref{table: risk factors} and these factors are grouped into ten aggregate risk classes based on type.

The degree distribution could be based directly on the reported numbers of P6M sexual contacts from the data. 
However, upon discussion we decided that this is likely to be underestimate the degrees due to the following reasons:
\begin{itemize}
    \item Stigma is widely recognized as a significant barrier to STBBI prevention, diagnosis and treatment~\cite{MacLean:2018}.
    Even those who provided information on their contacts might not be willing to reveal all of their contacts for fear of identification if public health notifies the contact of the risk of syphilis.
    
    \item Based on the high frequency of cases reporting \textit{anonymous sex}, \textit{sex trade} and \textit{judgement impaired due to alcohol or drugs} as risk factors, it is possible diagnosed individuals might not be able to remember or identify all of their contacts.

    \item The contacts listed within P6M may not reflect all the transmissible contacts.

    \item There were multiple individuals with risk factors usually associated with high numbers of sexual contacts (e.g. sex trade workers) who reported much lower numbers of contacts than would be expected based on the literature.
\end{itemize}

\begin{figure}[htbp]
    \begin{center}
        \includegraphics[width=0.9\textwidth]{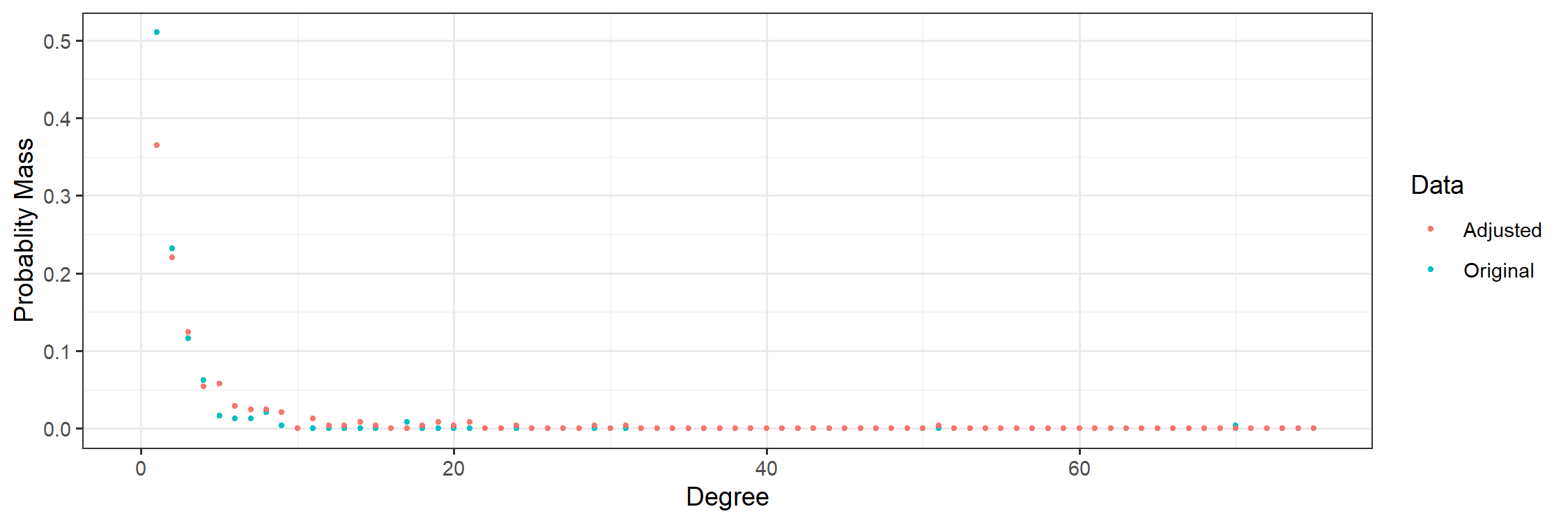}
    \end{center}
    \caption[Comparison of the original degree sequence and the adjusted degree sequence]{ Comparison of the original degree sequence and the adjusted degree sequence.
    The points represent the proportion of cases for each degree until $75$.}
    \label{fig: Data Adj}
\end{figure}

We adjusted up the reported number of sexual contacts of each case using their ages and reported risk factors.
The details of the adjustment are provided in Section~\ref{sec:rescaling_contacs}.
A comparison of the original degree sequence and the adjusted degree sequence is shown in Figure~\ref{fig: Data Adj}.

Studies based on empirical research \cite{LiljerosEtAl:2003, LiljerosEtAl:2001, AnneEtAl:2004} suggest that the degree distributions of sexual networks generally follow a power-law scale-free type of distribution. 
In such a distribution, for any randomly chosen vertex in the network, its degree $K$ equals to $k \in \mathbb{Z}_{+}$ with probability 
\begin{equation}
    \mathbb{P}(K=k) \propto k^{-\alpha}
    \label{eqn: power law}
\end{equation} 
for some constant $\alpha \in \mathbb{R}_{+}$.
We refer to this type of distribution as a power-law distribution.
Networks with this distribution are famous for its ``small-world''~\cite{WattsStrogatz:1998} and ``the rich get richer''~\cite{LiljerosEtAl:2001} effect, which are the small average path lengths between nodes and small amount of vertices have high degree while the majority of vertices have low degree.
This characteristic is also evident in our observations from the collected data as in Figure~\ref{fig: Data Adj}: more than $58.5\%$ cases would have degree less than $2$ even after modification adjustment, while $2$ cases have degree larger than $50$.
Thus, we take the following assumption for the degree distribution:
\begin{enumerate}
    \item [(H4)] The degree distribution of the target population follows a power-law distribution.
\end{enumerate}

\begin{figure}[htbp]
    \begin{center}
        \includegraphics[width=0.9\textwidth]{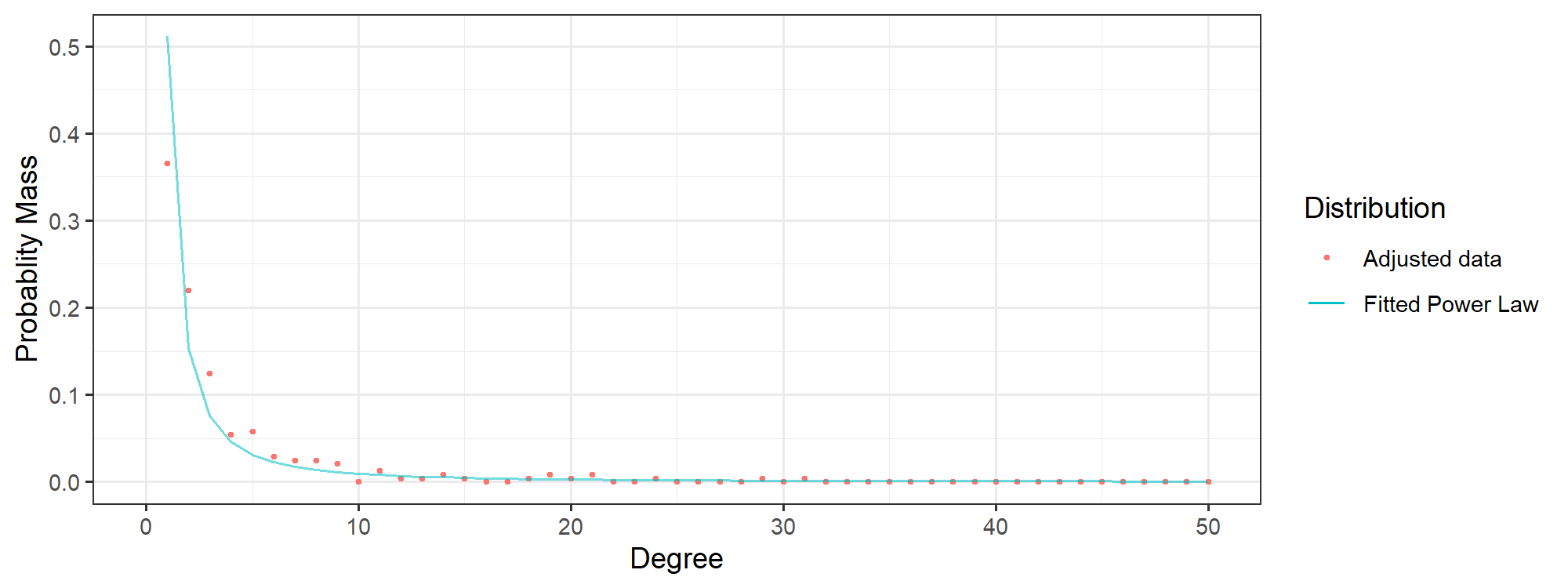}
    \end{center}
    \caption[Comparison between the adjusted sequence and fitted distribution]{ 
    For better illustration, the figure only shows probability from degree 1 to degree 50.}
    \label{fig: Degree Fit}
\end{figure}

We fitted the adjusted degree sequence to a power-law distribution using the \verb#fit_power_law# function from the \verb#igraph# R package, following the algorithm developed by Clauset, Shalizi and Newman~\cite{ClausetShaliziNewman_2009}.
The fitting resulted in a discrete power-law distribution with $\hat{\alpha}=1.738004$, which was also capped at maximum degree $200$ and normalized.
We denote $p_k=\mathbb{P}(K=k)$ for $k \in \{1,2,3,\cdots,200\}$ as the normalized probability mass function (PMF) for the degree distribution, satisfying \eqref{eqn: power law}.
The PMF of the fitted distribution is presented in Figure~\ref{fig: Degree Fit} together with the frequency probability distribution of the adjusted data sequence used for fitting.

\subsection{Epidemiological Assumptions}
\label{sec: Epi Dyn}
Individual differences on epidemiology parameters, like recovery rate and per-infection transmission rate, can be averaged out since the percolation model considers accounts for randomness in networks (see Zhao and Magpantay~\cite{ZhaoMagpantay:2024} for details).
Thus we assume the following:
\begin{enumerate}
    \item[(H5)] There is no individual differences between vertices other than their degree.
\end{enumerate}
As a result, all the vertices in the network are homogeneous on disease transmission dynamics, so the network is a single layer when considering dynamics on the compartment level.
This allows us to use uniform constant parameters when setting up the dynamics of every vertex.
This makes our model simpler, and also makes the comparison clearer between network-SIR model and MA-SIR model, since the only difference between them now is the social heterogeneity.

There are methods in the literature that consider multiple risk groups with different transmission parameters with multi-layer networks, which have different types of vertices and edges.
Bansal and Meyers~\cite{BansalMeyers:2012} created a framework based on Newman's typical percolation model \cite{NewmanStrogatzWatts:2001, Newman:2002,Newman:2003} for partial immunity.
Other works considering other types of network models are reviewed by Kinsley et. al.~\cite{KinsleyEtal:2020}.
However, multi-layer networks require more parameters which we are unable to ascertain due to our limited data.

It is also possible to consider more complex models than SIR-type models using percolation methods, such as those allowing for reinfection.
However, due to data limitation, we use the SIR compartment model used by Miller, Slim and Volz in their original paper~\cite{MillerSlimVolz:2012}.
\begin{enumerate}
\item[(H6)] There is no reinfection. The disease is modeling using SIR compartments.
\end{enumerate}
As shown in Figure~\ref{fig: proportions} (c), a few cases of reinfection were actually observed in the data.
For now we treat each reported reinfection case as an independent new case. 
Considering reinfection and immunity will be a future topic if more data can be obtained.

Under assumptions (H1)--(H6), the model satisfies the conditions to applying a modified bond percolation process on a configuration model (CM) network based on an SIR compartmental model. 
Following Miller, Slim and Volz~\cite{MillerSlimVolz:2012}, the dynamics of the disease transmission can be described by the following system of ordinary differential equations (ODEs):
\begin{equation}
\label{eqn: CM ODE}
    \begin{cases}
        S(t) & = G_p(\theta(t)) \\
        I(t) & = 1-S(t)-R(t) \\
        \dot{R}(t) & = \gamma I(t) \\
        \dot{\theta}(t) & = -\beta \theta +\beta \frac{G'_p(\theta)}{G'_p(1)}+\gamma (1-\theta)
    \end{cases} 
\end{equation}
Similar to the normalized ODE system of the traditional MA-SIR model (reviewed later in Section~\ref{sec: MA compare}), $S$, $I$ and $R$ respectively represent proportions that are susceptible, infectious and recovered.
The parameter $\beta$ is the uniform per-infected transmission rate and $\gamma$ is the uniform per-infected recovery rate.
Miller Slim and Volz~\cite{MillerSlimVolz:2012} introduced $\theta(t)$ to be the probability that a randomly chosen edge of a randomly chosen vertex in the network has not transmitted the infection at time $t$.
The function $G_p$ is the probability generating function (PGF) of the degree distribution, which we derived in the previous section.
Further details can be found in literature\cite{MillerSlimVolz:2012,ZhaoMagpantay:2024}.

To fit this network-SIR model~\eqref{eqn: CM ODE}  to data, we require initial conditions.
Since data collection began in 2019, we set the initial time $t=0$ to correspond to 2019 January 1st and set parameter $I_0$ to be the initial number of infected individuals.
We assume that initially the rest of the population is susceptible to the disease. 
The remaining parameters associated with \eqref{eqn: CM ODE} are the transmission rate $\beta$ and the recovery rate $\gamma$. 
In order to fit this model to data, we need to also account for imperfect reporting of syphilis, so we also introduce a reporting probability $p$ which is further discussed in the next section.
Thus the model has four epidemiological parameters ($I_0$ , $\beta$, $\gamma$ and $p$).

Before discussing the estimation of parameters (Sections~\ref{sec: reporting}--\ref{sec: MLE}), we note how the case data is related to the model trajectories.
Since the development of tertiary syphilis requires a long time (typically more than five years), the time frame we consider precludes tertiary cases from developing.
Similarly, we also exclude cases with unknown stage when fitting the dynamic of current outbreak. 
We aggregated the reported case numbers on a bi-weekly basis (14 days) when fitting the parameters.
Starting from 2019 January 1st, there are 116 bi-weeks until 2023 May 29th (the end time of data to be fitted), and 131 bi-weeks until 2024 January 9th (the end time of all the data available to us).
This leaves 230 reported cases that fall within the first 116 bi-weeks, including reinfection cases, which are treated as independent cases due to (H6).

From Figure~\ref{fig: Time Frame}, we also note that syphilis takes weeks and even months to go through each of its stages, so there is potentially a longer duration between exposure and being reported compared to the duration between reporting and treatment.
Thus, we consider reported cases as being reported as they transition from infectious to ``recovered'' instead of classifying them as actively infectious.
Additionally, we note that the data only contains the encounter date, where the case has been recorded and tested.
Before implementation of POCT, the only testing procedure for syphilis in KFL\&A area was the serology test through the laboratory of Public Health Ontario.
According to KFL\&A PH, the typical turn-around time (TAT) between getting tested and receiving treatment is 9 days, considering transportation, notification and a 4--5 days at laboratory.
Only after the test result is received will the individual get notification and then treatment.
To reflect this, we add 9 days to each encounter date of cases as our final date of recovery.

\section{Relationship Between Reporting Probability and Recovery Rate}
\label{sec: reporting}

In this section we show that the recovery rate $\gamma$ and reporting probability $p$ are not independent, and we present a simple way to tie them together.
As discussed in Section~\ref{sec: Syphilis}, syphilis is a treatable but not self-limiting disease.
Without treatment, syphilis will not self-recover, however individuals will move on to late stages that might not have noticeable symptoms for up to 20 years~\cite{PHACSyphilis}.
Syphilis is typically considered not to be transmissible when it develops to late latent stage or later, which typically happens after 12 months (26 bi-weeks) of exposure.

From the transmission dynamics point of view, we can consider an infected individual as ``recovered'' if they are no longer infectious, and thus if they have been treated or if they were not treated but are no longer transmitting the disease.
\begin{enumerate}
    \item[(H7)] The infected individual moves from infectious compartment to recovered compartment if one of the following happens:
    \begin{enumerate}
        \item[(a)] The individual is tested, treated and reported before their syphilis infection reaches late latent stage.

        \item[(b)] The individual's syphilis develops to late latent stage, regardless of whether they were tested, treated and reported afterwards.
    \end{enumerate}
\end{enumerate}
For simplicity, we assume a uniform constant probability of reporting:
\begin{enumerate}
    \item[(H8)] Every ``recovered'' individual has a probability $p$ of being tested, reported and recorded, regardless of time of testing and duration of infection.
\end{enumerate}

Now, recall that we aggregated the data by bi-week.
To fit the data, we also separate the contiguous time into short time intervals each with length of one bi-week by discrete time points $t \in \{0,1,2,\cdots,116\}$.
Consider the number of cases ``recovered'' during the time point $t$ to $t+1$ is given by $X(t)=(R(t+1)-R(t))N$. 
If $X(t)$ is an integer then since each individual has the same probability $p$ of reporting, the number of reported cases would follow a binomial distribution. 
We make the following assumptions:
\begin{enumerate}
\item[(H9)]  The number of reported cases $\hat{X}(t)$ is a random variable following the binomial distribution.

\item[(H10)] The reporting process of each time interval and the corresponding random variables $\hat{X}(t)$ are mutually independent for any $t$.
\end{enumerate}
The corresponding PMF is given by:
\begin{equation}
\label{eqn: binom}
    \mathbb{P}(\hat{X}(t)= n)= f_{binom}(X(t),n,p)=\binom{X}{n} p^n (1-p)^{X(t)-n} \text{, } \forall n \in \{n\in \mathbb{Z}_{+}|n \leq X(t)\}
\end{equation}
In the next section we will introduce the continuous binomial distribution adjustment used when $X(t)$ is not an integer.
For now, we note that $E( \hat{X}(t))=X(t) \times p$.
By (H9), the expectation of total reported recovered case until any time point $T$ is given by, 
\begin{equation}
\label{eqn: report}
    \sum_{t=0}^{T-1} \mathbb{E} (X(t))= \sum_{t=0}^{T-1} X(t)p = p \sum_{t=0}^{T-1} N(R(t+1)-R(t))=R(T)Np.
\end{equation}

As discussed in Zhao and Magpantay~\cite{ZhaoMagpantay:2024}, the application of the system \eqref{eqn: CM ODE} also requires the assumption that the infectious period (or ``recovery time'') of all infected cases follows an exponential distribution with rate $\gamma$.
The base MA-SIR model also requires similar assumptions.
For such an assumption, the expectation of the infectious period is equal to the infectious duration given by $1/\gamma$.
The exponential distribution, together with (H7) and (H8) indicates that we can find a positive correlation between $p$ and $\gamma$.
Increasing reporting probability $p$ means more cases are reported and treated.
Since we assume $p$ is the same for all different stages, this results in more cases being reported and getting treatment at early stages, thus lowering the average infectious period and leading to higher $\gamma$.

As shown in Figure~\ref{fig: proportions}, the collected data includes stage of syphilis based on interpretation of the serological test results.
Therefore, we can estimate the infectious duration of reported cases, which we can use to further simplify the model by creating a bijection between $p$ and $\gamma$.

\begin{figure}[htbp]
    \begin{center}
        \includegraphics[width=0.8\textwidth]{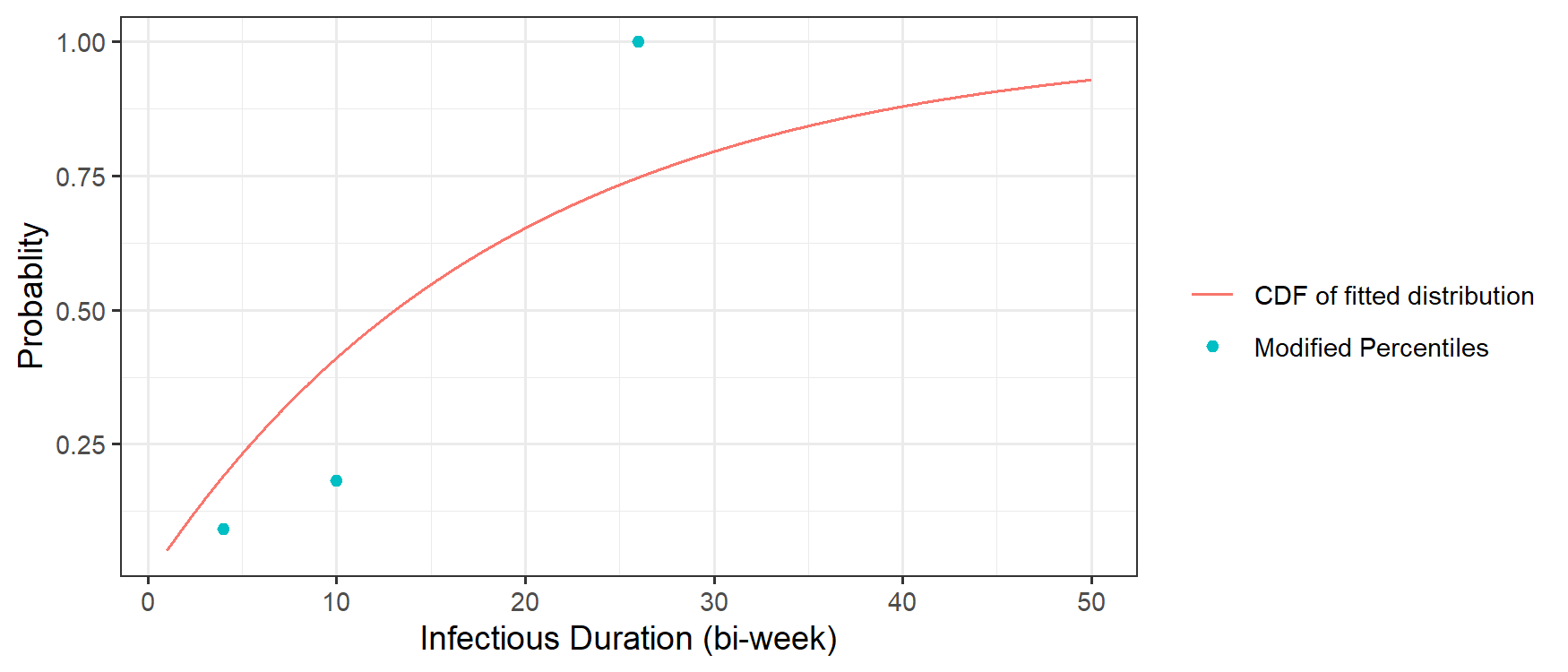}
    \end{center}
    \vspace{-5mm}
    \caption[Example of fitting the exponential distribution from percentiles to determine ``recovery'' rate $\gamma$ from reporting probability $p$]{Example of fitting the exponential distribution from percentiles to determine ``recovery'' rate $\gamma$ from reporting probability $p$. Here we set $p=0.3$ as input. The resulting estimate is $\hat{\gamma}=0.5286975$.}
    \label{fig: Per Fit}
\end{figure}

Consider the 292 total reported cases and \eqref{eqn: report} based on the properties of independent binomial distributions, we take $R^{*}=\frac{292}{p}$ as an estimate of the total recovered number $R(T)N$.
We now use this on the data to find $\gamma$ as a function of $p$.
With (H7), the cases reported at primary, secondary and early latent stage would be the only cases that recovered before natural loss of transmission ability at the late latent stage.
Based on typical duration of each syphilis stage given in Figure~\ref{fig: Time Frame}, there are 88 reported cases of primary syphilis so we made an initial estimate that $P_1=88/R^{*}$ is the proportion of all ``recovered'' cases that recovered before four bi-weeks (eight weeks is the estimated typical duration of primary syphilis).
Next we noted that there are $89$ reported cases of secondary syphilis so we made an initial estimation $P_2=P_1+89/R^{*}$ as the proportion of ``recovered'' cases before ten bi-weeks (20 weeks is the estimated end of secondary syphilis).
Since neurosyphilis can occur at any stage, for consistency and simplicity, we distributed all 9 cases uniformly across 130 bi-weeks (about 5 years) so that each bi-week duration gets $\frac{9}{130}$ value of adjustment.
As a result of adjusting we get the probabilities of ``recovery' at four and ten bi-weeks respectively to be $P_1^{*}=P_1+4 \times \frac{9}{130}$ and $P_2^{*}=P_2+10 \times \frac{9}{130}$.
By (H7), we assume that the infectious period in most cases is finished before the late latent stage so we add $P_3^*\approx 100\%$ at $t=26$ bi-weeks.
We fitted these three percentiles (at four, ten and 26 bi-weeks) with an exponential distribution using the ``get.exp.par'' function from the R package \verb#rriskDistributions#, which finds the optimized exponential distribution with least sum of square difference between the given probabilities and the theoretical probabilities evaluated at the given percentile points.
An example of such optimization result with input percentiles is shown in Figure~\ref{fig: Per Fit}.

\begin{figure}[htbp]
    \begin{center}
        \includegraphics[width=0.6\textwidth]{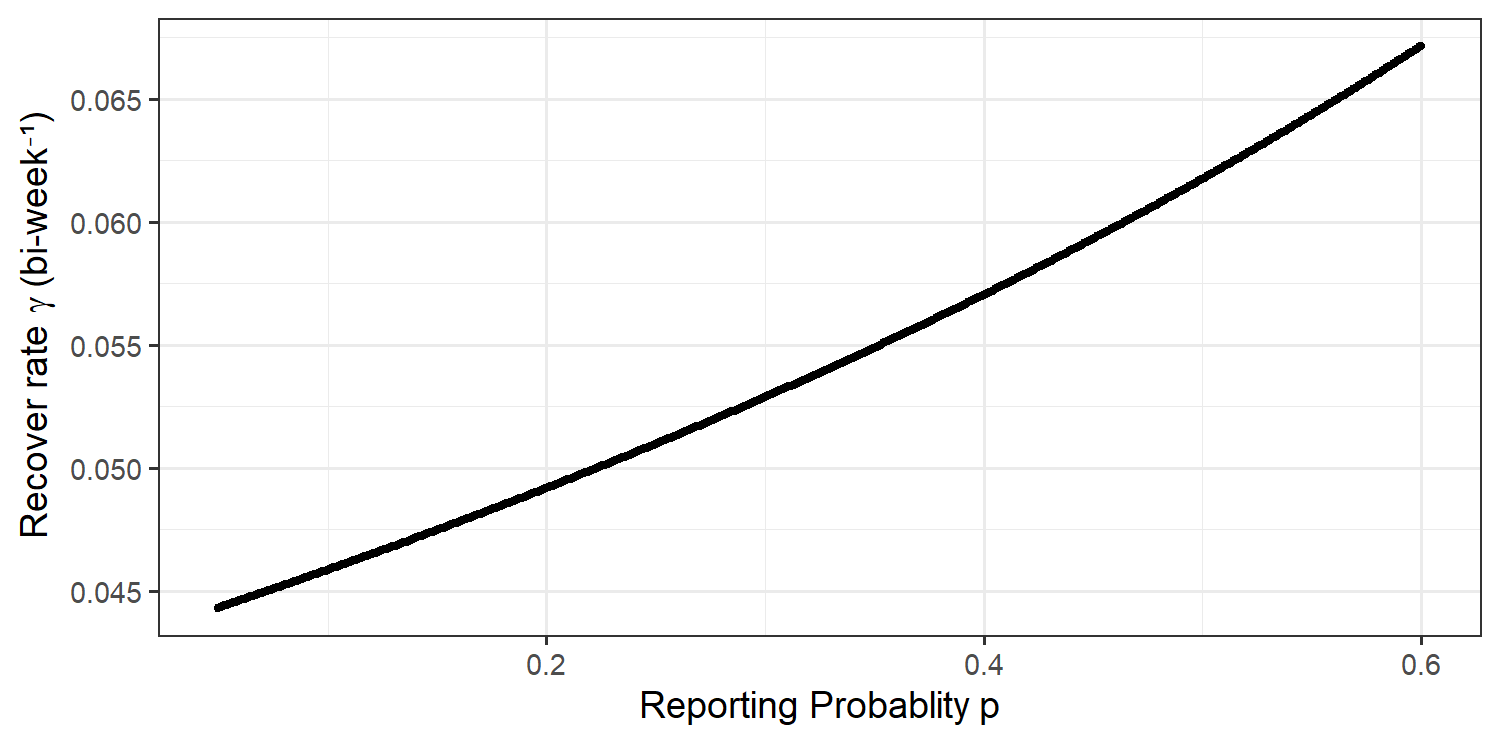}
    \end{center}
    \vspace{-5mm}
    \caption[Relationship between ``recovery'' rate $\gamma$ and reporting probability $p$]{Relationship between ``recovery'' rate $\gamma$ and reporting probability $p$, following the  fitting shown in Figure~\ref{fig: Per Fit}.}
    \label{fig: Gamma p}
\end{figure}

The departure from the percentiles is also partially offset by the individual difference in ``recovery.''
The only variable during such fitting procedure is $p$, and it generates a unique $\gamma$ value for each given $p$, which reflect the correlation as shown in Figure~\ref{fig: Gamma p}

\section{Maximum Likelihood Estimation of Parameters}
\label{sec: MLE}

We showed in the previous section that the recovery rate $\gamma$ can be determined from a given value of $p$.
Thus, to fit the model to the time series data we only need to find estimations of three parameters: the initial number of infectious individuals $I_0$, the reporting probability $p$ and the transmission rate $\beta$.
We did this using maximum likelihood.

\subsection{Computation of Likelihood}
\label{sec: Lik Exp}

Here we describe how to compute the likelihood that a given set of values of the parameters $I_0$, $p$ and $\beta$ gave rise to our time series of syphilis cases aggregated by bi-weeks.
We set $I(0)=\frac{I_0}{N}$ (where $t=0$ corresponds to 1 January 2019). 
Since, there were limited cases reported before the initial time then we also assume that the start of this outbreak is not likely to be more than one year prior to 2019.
Thus we set $R(0)=0$ since there should not be enough time for any infected cases develop to late latent stage.
Therefore, we have the following initial conditions: $I(0)=\frac{I_0}{N}$, $R(0)=0$, $S(0)=1-I(0)-R(0)=1-\frac{I_0}{N}$ and $\theta(0)$ can be found as solution of the equation
\begin{equation}
    1-\frac{I_0}{N}=G_p(\theta(0)).
\end{equation}
Solving this yields all the required initial conditions to generate the trajectory of \eqref{eqn: CM ODE} for a given $I_0$, $p$ and $\beta$.

\begin{figure}[htbp]
    \begin{center}
        \includegraphics[width=0.6\textwidth]{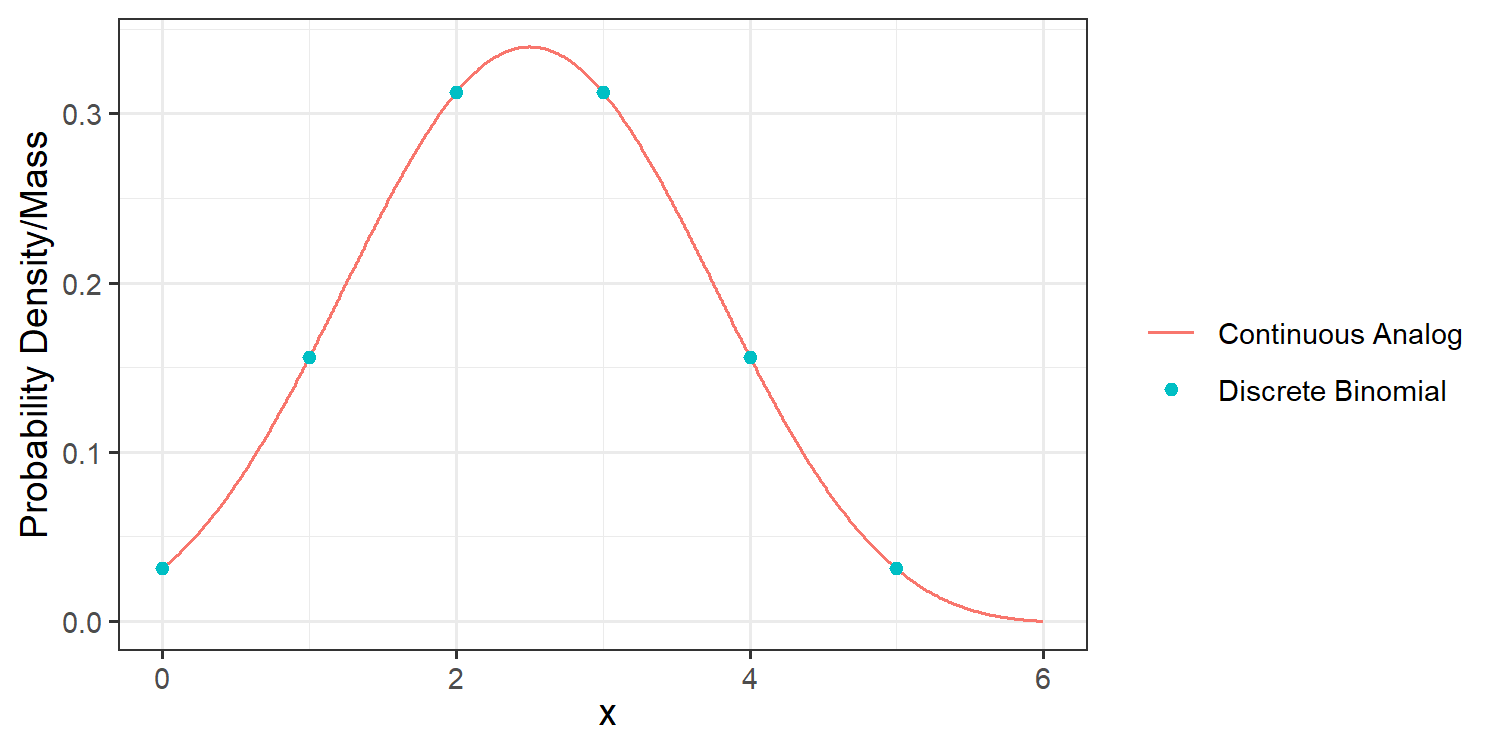}
    \end{center}
    \vspace{-5mm}
    \caption[Continuous Analogue of the Binomial Distribution]{Comparison between the PMF of binomial distribution and the PDF of its continuous analog~\eqref{eqn: CB PDF} with $X=5$ and $p=0.5$.}
    \label{fig: CB}
\end{figure}

Given any trajectory of the recovered individuals, (whether computed via an MA-SIR model or the network-SIR model) we can determine the likelihood using the same under-reporting binomial model discussed in Section~\ref{sec: reporting}.
We use the notation $R(t;\vec{\lambda})$ for the $R(t)$ trajectory given parameter vector $\vec{\lambda}=(p,\beta, I_0)$.
Therefore, the number of recoveries over each time duration between $t$ and $t+1$ (one bi-week later) is denoted by $X(t;\vec{\lambda})=N\big[R(t+1;\vec{\lambda})-R(t;\vec{\lambda})\big]$.
Since $X(t)$ is not necessarily an integer and we are unable to use the standard discrete binomial distribution as described in \eqref{eqn: report}.
Instead, to compute the likelihood we consider a continuous analogue of the binomial distribution~\cite{Ilienko:2013}, which has cumulative density function given by,
\begin{equation}
    \label{eqn: CB CDF}
    F_{CB}(X,n,p)=
    \begin{cases}
        0, & n \in (-\infty,0], \\
        \frac{B(n,X+1-n,p)}{B(k,X+1-n,0)}, & n \in (0,X+1],\\
        1, & n \in (X+1, +\infty).\\
    \end{cases}
\end{equation}
Here $B$ is the incomplete $\beta$ function given by,
\begin{equation}
    \label{eqn: bfunc}
    B(x,y,p)= \int_{p}^{1} t^{X-1} (1-t)^{y-1} dt \text{ , } x,y>0, p\in [0,1].
\end{equation}
For our computation of likelihood, we approximate the probability density function by,
\begin{equation}
    \label{eqn: CB PDF}
    f_{CB}(X,n,p) \approx \hat{f}_{CB}(X,n+1,p) = F_{CB} (X,n+1,p)-F_{CB}(X,n,p).
\end{equation}
We note that unlike the discrete binomial distribution which requires $n \leq X$, the continuous analog allows $X < n \leq X+1$.
Thus to compute the likelihood, we replace the $f_{binom}$ in \eqref{eqn: binom} by $\hat{f}_{CB}$ in \eqref{eqn: CB PDF} when fitting the data.
Figure~\ref{fig: CB} provides an example comparing the $f_{binom}$ with $\hat{f}_{CB}$.

Now let $\hat{n}(t)$ be the adjusted aggregated bi-weekly reported case count from the data, where $t\in \mathbb{Z}_{+}\cap[0,116)$, and let $\Vec{\hat{n}}=(\hat{n}(0),\hat{n}(1),\cdots,\hat{n}(115))$ be the vector of data.
Thus, given case count data $\vec \hat n$ for the first 116 bi-weeks, for any parameter vector $\vec{\lambda}=(p,\beta,I_0)$ using the binomial under-reporting model the log-likelihood function is given by,
\begin{equation}
    \label{eqn: loglik}
    \ell(\vec{\hat n}; \Vec{\lambda})=\sum_{t=0} ^{115} \log(\hat{f}_{CB} (X(t;\Vec{\lambda}),\hat{n}(t),p)) 
\end{equation}

\subsection{Maximum Likelihood Estimates}
\label{sec: result}

Using \eqref{eqn: loglik} we searched for the maximum likelihood estimate (MLE) of the parameters $I_0$, $\beta$ and $p$ for the network-SIR model. 
The result is presented in Table~\ref{table: MSV Init27}.
All model trajectories were generated by \verb#EpiNetPerco#, the R package that we developed.
Code used to generate the figures in this paper will be uploaded as part of \verb#EpiNetPerco# documentation.

\subsection*{Finding the optimal $I_0$}

\begin{figure}[htbp]
    \begin{center}
        \includegraphics[width=0.6\textwidth]{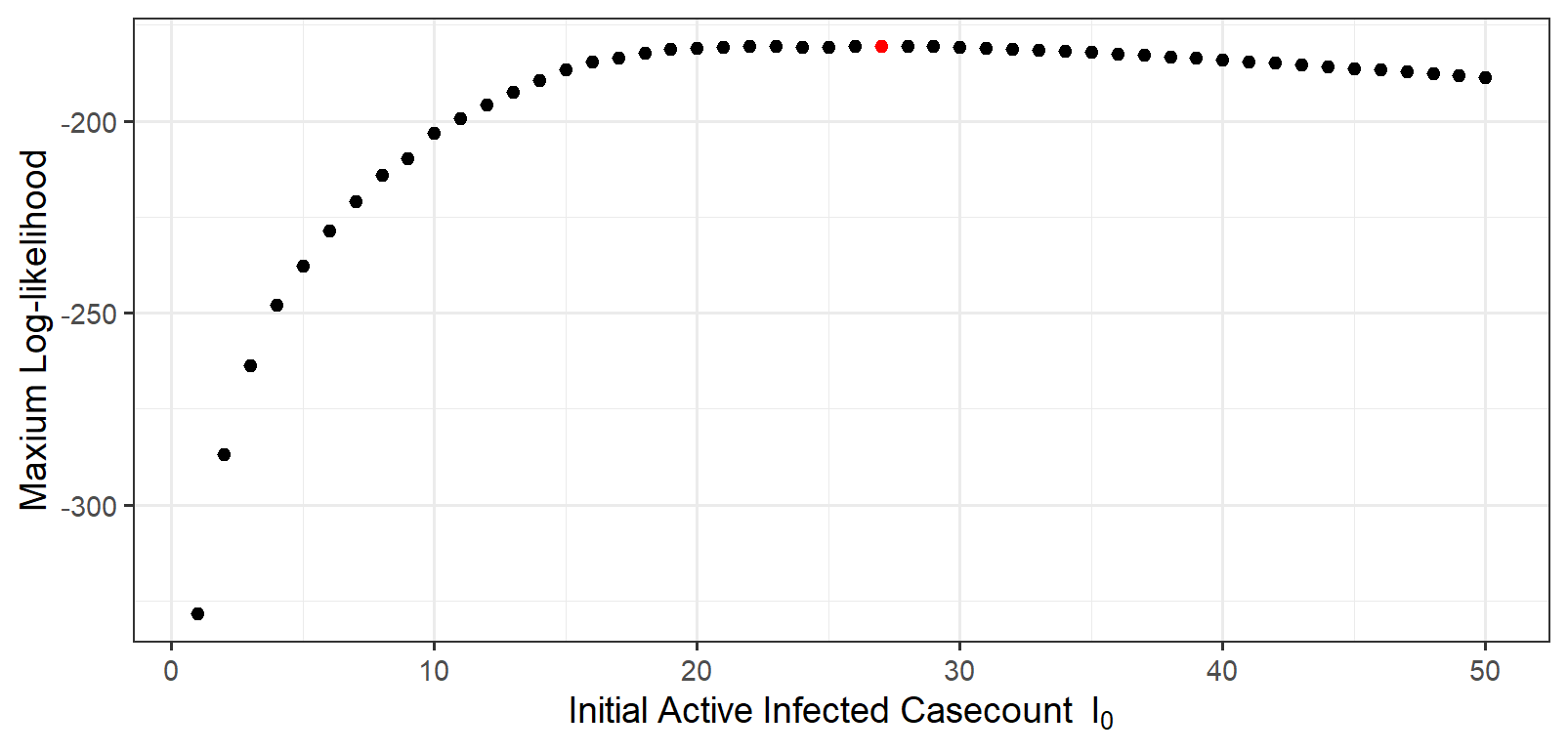}
    \end{center}
    \vspace{-5mm}
    \caption[Profile over $I_0$ for the network-SIR model]{Profile over $I_0$ for the network-SIR model. The red point corresponds to the  optimal parameter value $\hat{I}_0 = 27$.}
    \label{fig: MSV I0}
\end{figure}

To better illustrate the results, we first look at the likelihood profile over $I_0$.
For each fixed $I_0$ value, we calculate the log-likelihood for a large grid of values of the other parameters $\beta$ and $p$ (for example as in Figure~\ref{fig: MSV Init27}), and recorded the maximum log-likelihood attained.
This plot is shown in Figure~\ref{fig: MSV I0}. 
We see there also that $\hat{I}_0=27$ attains the highest likelihood. 

\subsection*{Log-likelihood as a function of $p$ and $\beta$ for the  optimal $I_0$}

\begin{figure}[htbp]
    \centering
    \begin{subfigure}[b]{0.6\textwidth}
         \centering
         \includegraphics[width=\textwidth]{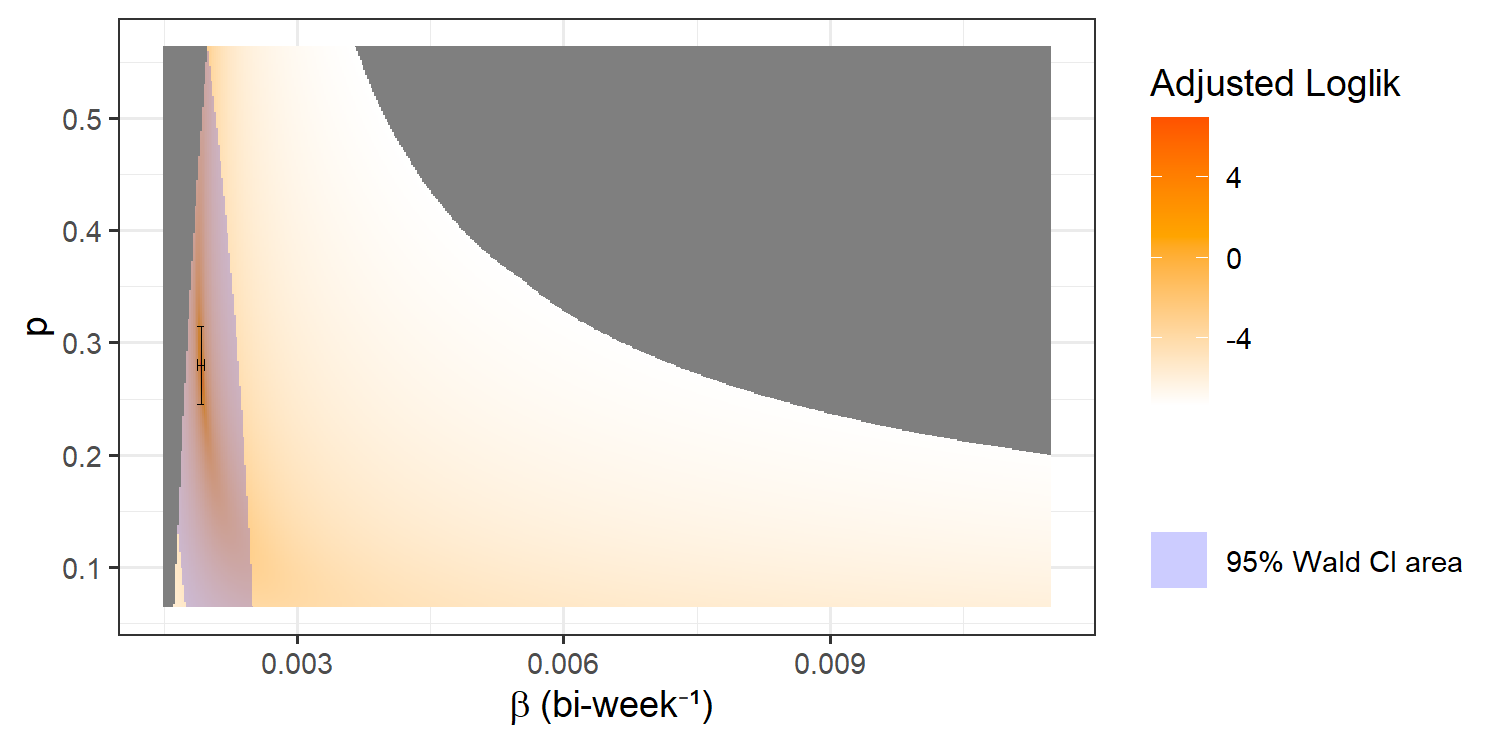}
         \caption{Heat map with Wald 95\% CIs}
    \end{subfigure}
    \begin{subfigure}[b]{0.38\textwidth}
         \centering
         \includegraphics[width=\textwidth]{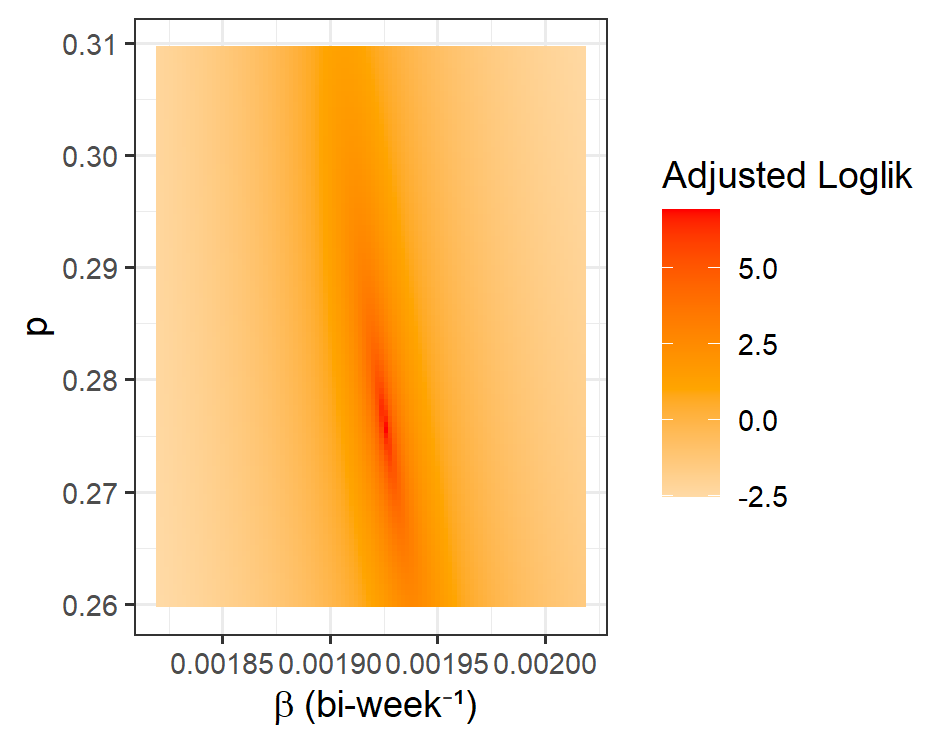}
         \caption{Neighbourhood of the maximum likelihood}
    \end{subfigure}
    \caption{Heat map of log-likelihood for the network-SIR model with $\hat{I}_0=27$. In (a) the single variable confidence interval is shown using the black bars while the multivariate confidence area is shaded in blue.}
    \label{fig: MSV Init27}
\end{figure}

The log-likelihood as a function of $p$ and $\beta$ is shown as a heat map in Figure~\ref{fig: MSV Init27}.
For clarity, the color in the heat map is actually determined by the adjusted log-likelihood $h$ which is given by:
\begin{equation}
    h=h(\hat{I}_0; \beta, p)=- \log ( \ell(\Vec{n}; \hat{I}_0, \hat{\beta}, \hat{p}) - \ell(\Vec{n}; \hat{I}_0, \beta, p) +0.001).
\end{equation}
This means the higher $h$ is, the closer it is to red in the heat map and the corresponding log-likelihood is higher and closer to the maximum likelihood.
This adjustment is applied to all the heat maps in this paper.

In Figure~\ref{fig: MSV Init27} (a), the grey area indicates parameters values for which some data points $n(t)$ are actually larger than $X(t)+1$ from the prediction, which gives it a value of $\hat{f}_{CB}=0$ and corresponding log-likelihood is $\ell=-\infty$.
We also present the Wald 95\% confidence interval for each single variable and the multivariate confidence area.

\begin{figure}[htbp]
    \centering
    \begin{subfigure}[h]{0.9\textwidth}
         \centering
         \includegraphics[width=0.9\textwidth]{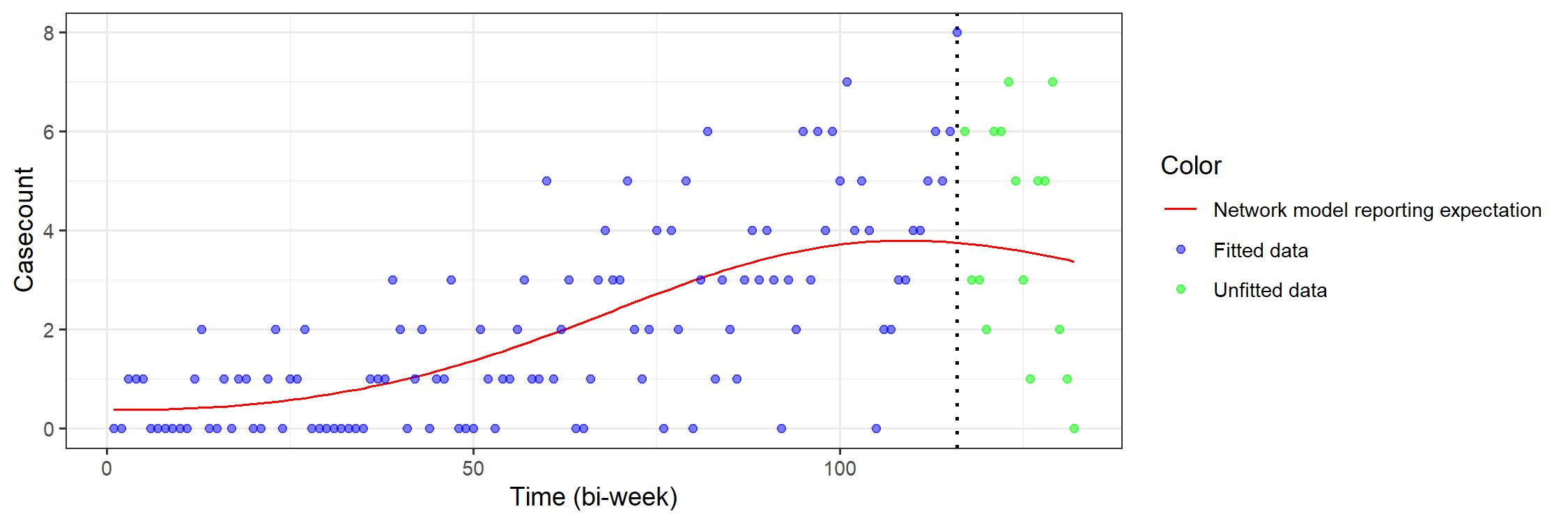}
         \caption[MSV R incident]{Bi-weekly reported recovered cases}
    \end{subfigure}
    \begin{subfigure}[h]{0.9\textwidth}
         \centering
         \includegraphics[width=0.9\textwidth]{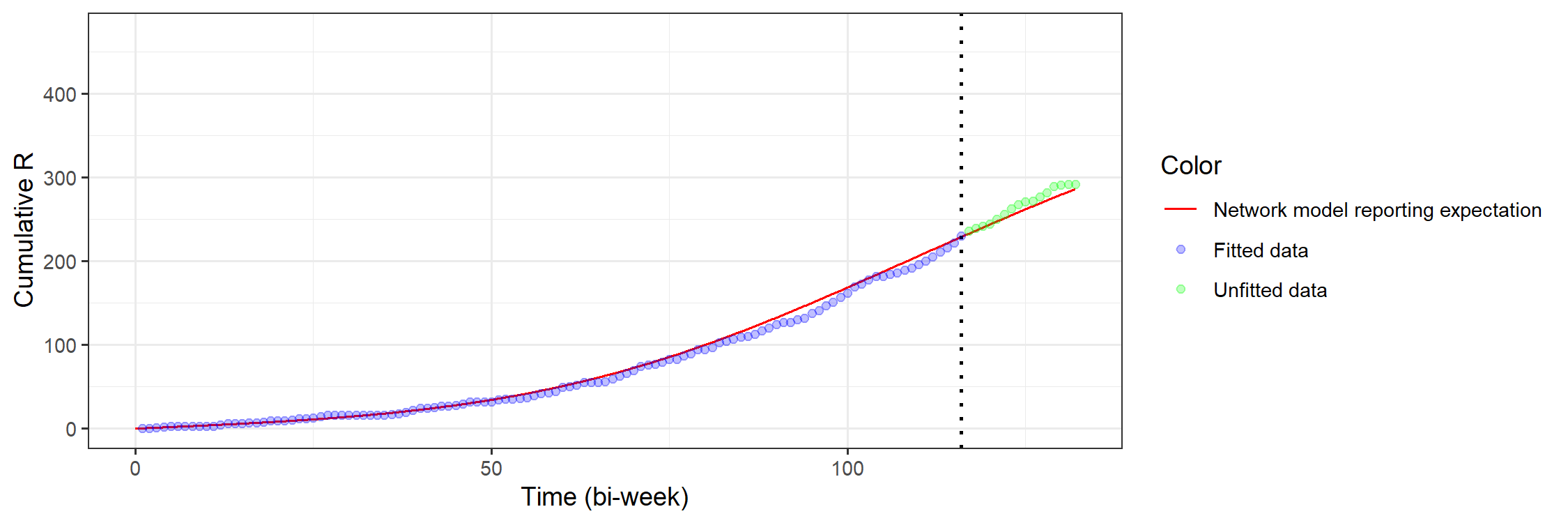}
         \caption[MSV R Cumulative]{Cumulative reported recovered cases}
    \end{subfigure}
    \caption{Comparison of the trajectories of the expected reported cases of the network-SIR model with data. The dotted vertical lines in both sub-figure reflect the date when the
POCT was implemented.}
    \label{fig: MSV R Cumulative}
\end{figure}

The results are presented in Table~\ref{table: MSV Init27}.
We included here the epidemiological predictions, such as the final infected proportion and final infected size, computed using our R package \verb#EpiNetPerco#. 
Further details on the theory behind this are provided in Zhao and Magpantay~\cite{ZhaoMagpantay:2024}.
We also present a comparison of the expected trajectory of \eqref{eqn: report} and all of aggregated data in Figure~\ref{fig: MSV R Cumulative}.

\vspace*{-5mm}
\begin{center}
    \begin{table}[htb]
    \centering
        \caption{MLEs and predictions of the network-SIR model. Rates ($\hat{\gamma}$ and $\hat{\beta}$) have unit biweek$^{-1}$.
        The intervals in bracket are the $95\%$ Wald CI of corresponding fitted single parameter.}
    \label{table: MSV Init27}
        \begin{tabular}{|C{0.6in}|C{2.8in}|C{1.2in}|}
            \hline
            Notation & Definition & Model Value
            \\
            \hline
            $N$ & Population/Network Size & 26,000
            \\
            \hline
            $\hat{\alpha}$ & Optimal power law parameter of degree distribution & 1.738004 
            \\
            \hline
            $\hat{I}_0$ & Optimal initial active case count & 27
            \\
            \hline
            $\hat{\beta}$ & Optimal uniform per-infected transmission rate & 0.001926  $(0.001878, 0.001974)$
            \\
            \hline
            $\hat{p}$ & Optimal uniform reporting probability & 0.2755 $(0.2412, 0.3098)$ 
            \\
            \hline
            $\hat{\gamma}$ & Optimal uniform per-infected recovery rate & 0.05192518 
            \\
            \hline
            $\ell$ & Maximal log likelihood & -180.2353
            \\
            \hline
            $R_{0}$ & Basic reproductive number & 1.8548
            \\
            \hline
            $R(\infty)$ & Final infected proportion & 0.06606
            \\
            \hline
            $R(\infty) \times N$ & Final infected size & 1717.59
            \\
            \hline
        \end{tabular}
    \end{table}
\end{center}
\vspace{-6mm}

\section{Comparing the network-SIR and MA-SIR models}
\label{sec: MA compare}
To illustrate the difference between network-SIR and MA-SIR models, we present the results of fitting the standard MA-SIR model using the same data, same under-reporting model and same algorithms. 
The only thing we need to alter is the assumption about social structure.
Instead of network characterized by degree distribution discussed in Section~\ref{sec: Net&Deg}, we assume that the population is fully mixed and the dynamics are given by, 
\begin{equation}
\label{eqn: MA ODE}
    \begin{cases}
        \dot{S}(t) & = -\beta S(t) I(t) \\
        \dot{I}(t) & = \beta S(t) I(t)-\gamma I(t) \\
        \dot{R}(t) & = \gamma I(t) \\
    \end{cases} 
\end{equation}
We again assume $I(0)=I_0$ and the rest of the population is susceptible at time $t=0$.

\begin{figure}[p]
    \begin{center}
        \includegraphics[width=0.6\textwidth]{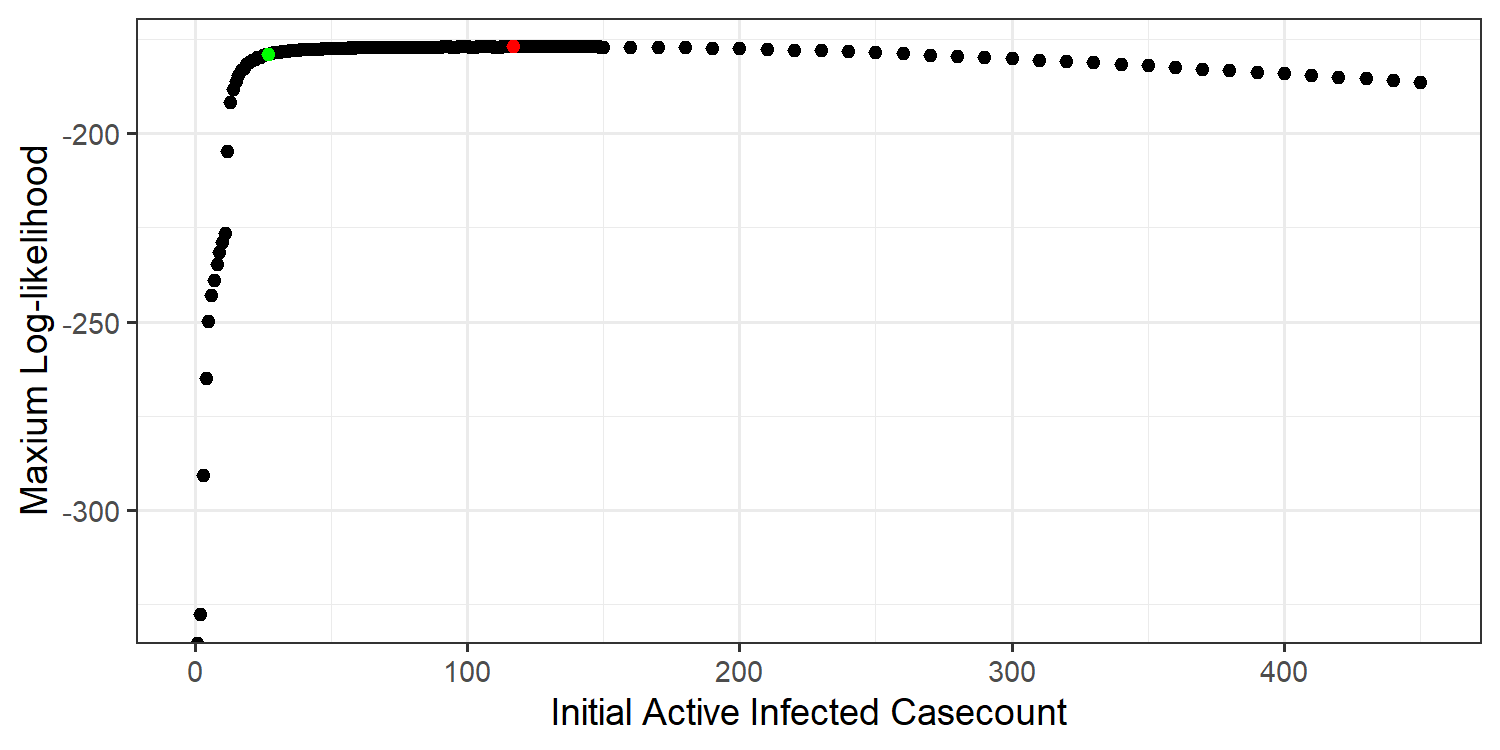}
    \end{center}
    \caption[MASIR $I_0$ Fitting]{Maximum log-likelihood of different $I_0$ value for MA model. The red point is the optimal value $\hat{I}_0 = 117$, and the green point is the optimal value for network based model $I_{0}=27$.}
    \label{fig: MASIR I0}
\end{figure}

In Figure~\ref{fig: MASIR I0} we present the profile likelihood over the initial condition $I_0$, which can be compared directly to the results we found for the network-SIR model in Figure~\ref{fig: MSV I0}. 
We see that the MA-SIR model is far less sensitive to the initial number of infected individuals due to tradeoffs with other parameters (such as the reporting probability). 
In this case, the likelihood profile is relatively flat for a large range of values and the MLE is at a much higher initial active case count at $\hat{I}_0 = 117$ compared to $27$ for the network-SIR model.
We present the results of MA-SIR model with both $I_0=27$ and $I_0=117$, and compare these with the results from the network-SIR model.

\begin{figure}[p]
    \centering
    \begin{subfigure}[b]{0.6\textwidth}
         \centering
         \includegraphics[width=\textwidth]{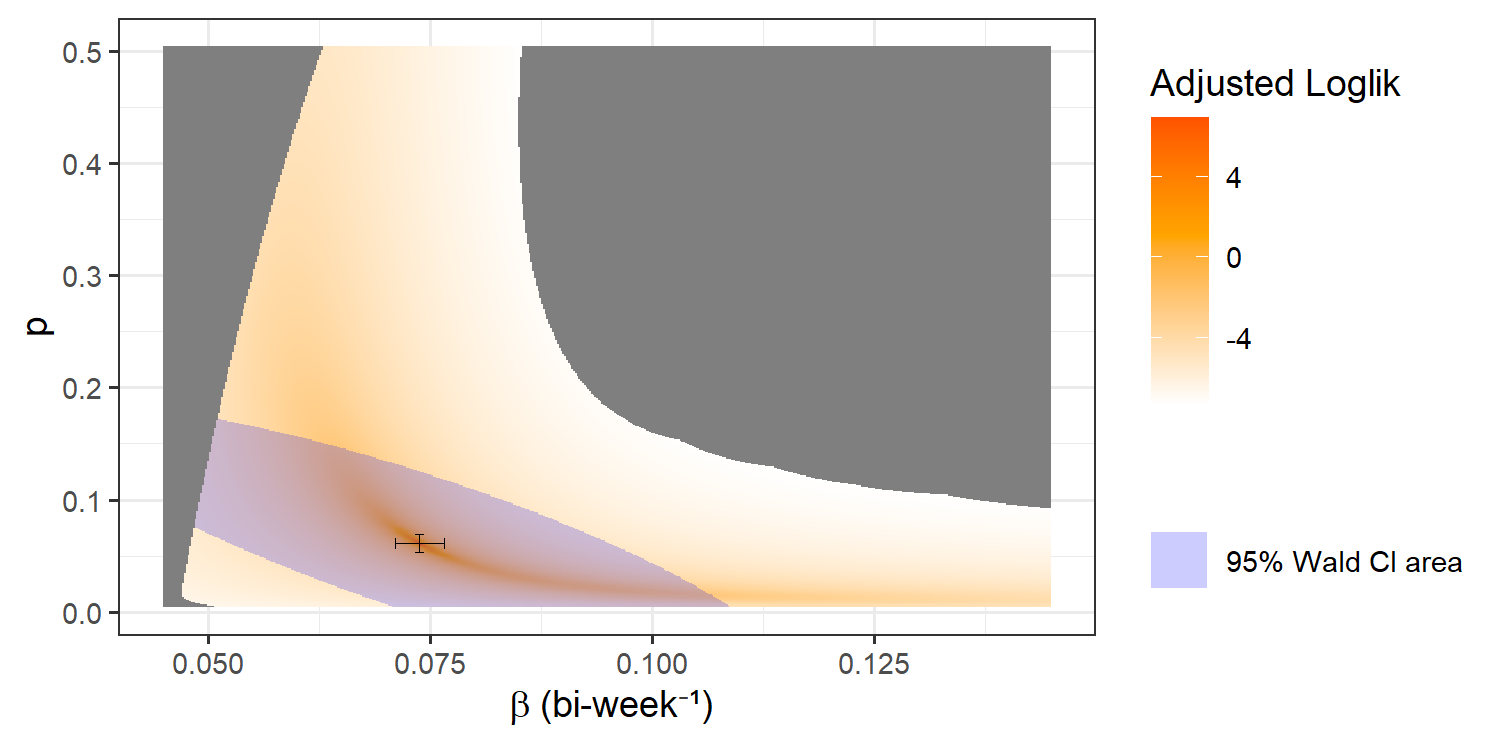}
         \caption{Heat map with Wald 95\% CIs}
    \end{subfigure}
    \begin{subfigure}[b]{0.38\textwidth}
         \centering
         \includegraphics[width=\textwidth]{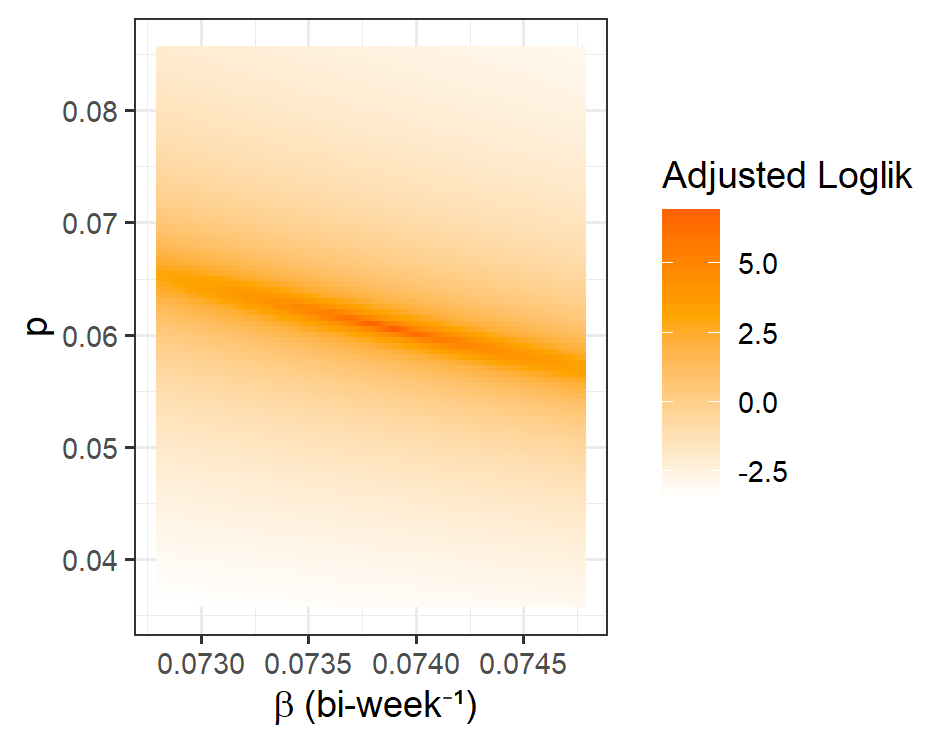}
         \caption{Neighbourhood of the maximum likelihood}
    \end{subfigure}
    \caption{Heat map of log-likelihood for the MA-SIR model with $\hat{I}_0=117$}
    \label{fig: MASIR Init117}
\end{figure}

\begin{figure}[p]
    \centering
    \begin{subfigure}[b]{0.6\textwidth}
         \centering
         \includegraphics[width=\textwidth]{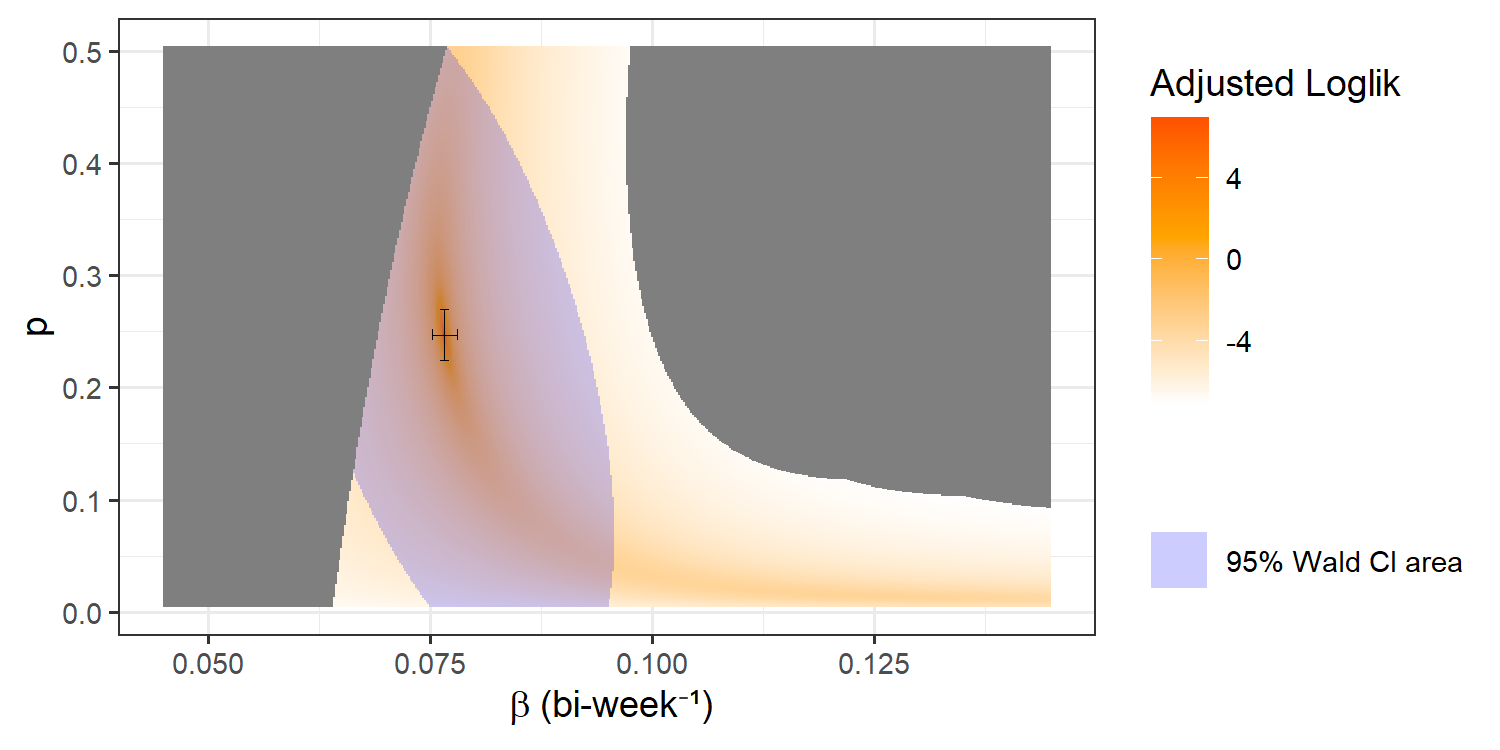}
         \caption{Heat map with Wald 95\% CIs}
    \end{subfigure}
    \begin{subfigure}[b]{0.38\textwidth}
         \centering
         \includegraphics[width=\textwidth]{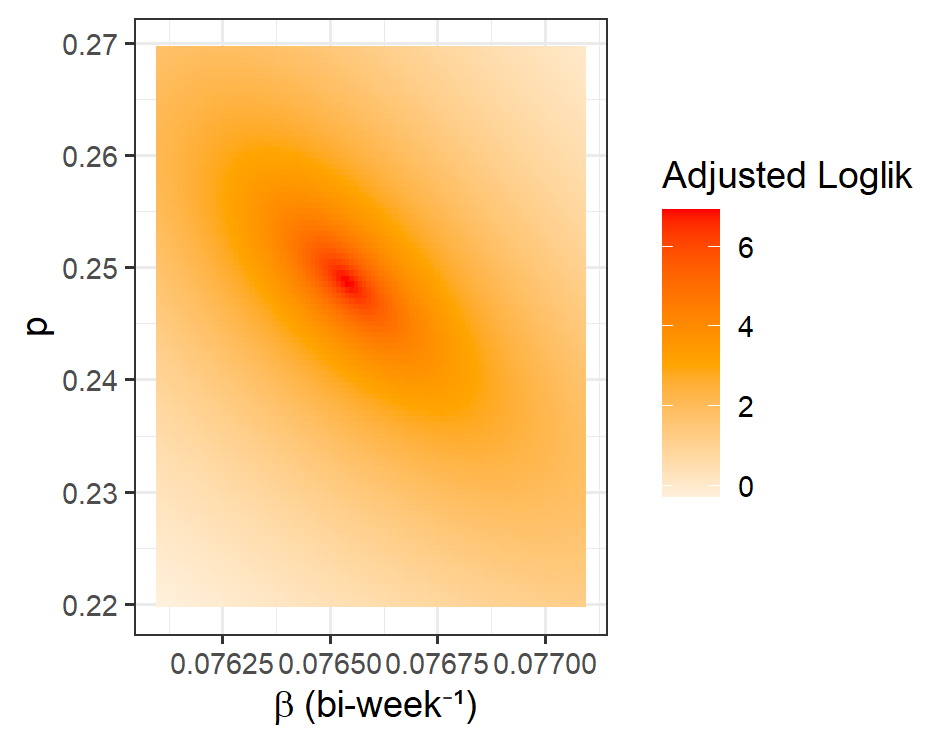}
         \caption{Neighbourhood of the maximum likelihood}
    \end{subfigure}
    \caption{Heat map of log-likelihood for the MA-SIR model with $I_0=27$}
    \label{fig: MASIR Init27}
\end{figure}

For the $\hat{I}_0=117$ case, the heat map of log-likelihood is shown in Figure~\ref{fig: MASIR Init117}, the numerical parameters, confidence intervals and predictions are given in Table~\ref{table: MASIR Init117}.
For the $I_0=27$ case, these are shown in Figure~\ref{fig: MASIR Init27} and Table~\ref{table: MASIR Init117}.

\vspace{-5mm}
\begin{center}
    \begin{table}[p]
    \centering
        \caption{MLE and predictions of the  MA-SIR model with $\hat{I}_0=117$ and $I_0=27$. 
        Rates ($\hat{\gamma}$ and $\hat{\beta}$) have unit biweek$^{-1}$. 
        The intervals in bracket are the $95\%$ Wald CI of corresponding fitted single parameter.}
    \label{table: MASIR Init117}
        \begin{tabular}{|C{0.6in}|C{2.8in}|C{1.2in}|C{1.2in}|}
            \hline
            Notation & Definition & MA-SIR $\hat{I}_{0}=117$ & MA-SIR $I_{0}=27$
            \\
            \hline
            $N$ & Population/Network Size & \multicolumn{2}{c|}{26,000}
            \\
            \hline
            $\hat{I}_0$ & Optimal initial active case count & 117 & N/A
            \\
            \hline
            $I_0$ & Assumption initial active case count & N/A & 27
            \\
            \hline
            $\hat{\beta}$ & Optimal uniform per-infected transmission rate & 0.007390 $(0.07089, 0.07691)$ & 0.07654 $(0.07513, 0.07795)$
            \\
            \hline
            $\hat{p}$ & Optimal uniform reporting probability & 0.0605 $(0.0516, 0.0694)$ & 0.2485 $(0.2251, 0.2719)$
            \\
            \hline
            $\hat{\gamma}$ & Optimal uniform per-infected recovery rate & 0.04463588 & 0.05091438
            \\
            \hline
            $\ell$ & Maximal log likelihood & -176.8698 & -178.8482
            \\
            \hline
            $R_{0}$ & Basic reproductive number & 1.6556 & 1.5033
            \\
            \hline
            $R(\infty)$ & Final infected proportion & 0.6731 & 0.5835
            \\
            \hline
            $R(\infty) \times N$ & Final infected size & 17501.32 & 15171.09
            \\
            \hline
        \end{tabular}

    \end{table}
\end{center}
\vspace{-6mm}

\begin{figure}[p]
    \centering
    \begin{subfigure}[b]{0.9\textwidth}
         \centering
         \includegraphics[width=0.9\textwidth]{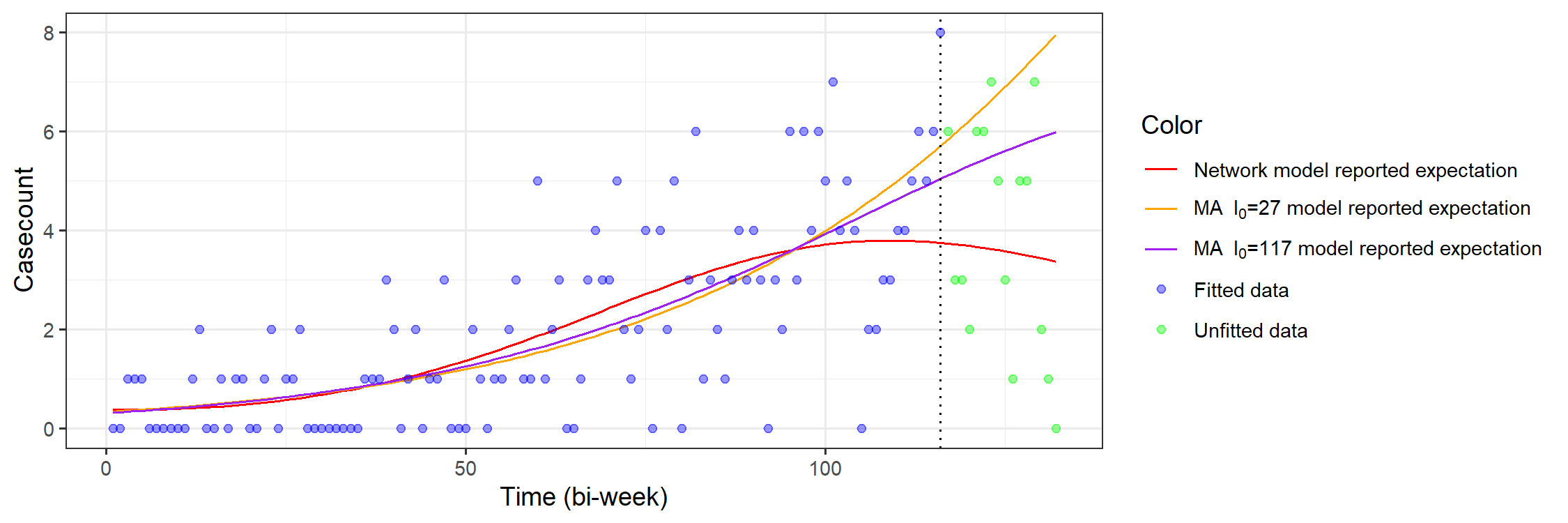}
         \caption[MASIR R incident]{Bi-weekly reported recovered cases.}
    \end{subfigure}
    \begin{subfigure}[b]{0.9\textwidth}
         \centering
         \includegraphics[width=0.9\textwidth]{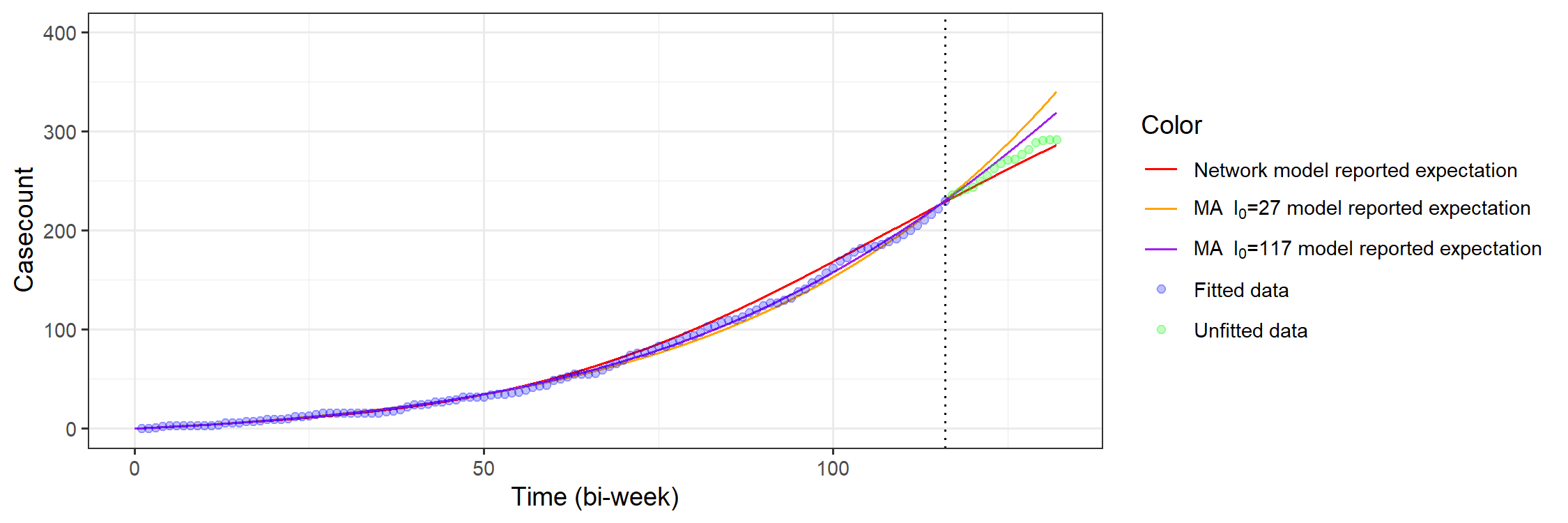}
         \caption[MSV R Cumulative]{Cumulative reported recovered cases.}
    \end{subfigure}
    \caption[Comparison of the trajectories of the expected reported cases generated by different models with data
    ]{Comparison of the trajectories of the expected reported cases of the two MA-SIR models and the network-SIR model with data.
    The dotted vertical lines in both sub-figure reflect the date when the POCT was implemented.}
    \label{fig: MASIR R Cumulative}
\end{figure}

\begin{figure}[h]
    \centering
    \begin{subfigure}[b]{0.49\textwidth}
         \centering
         \includegraphics[width=1\textwidth]{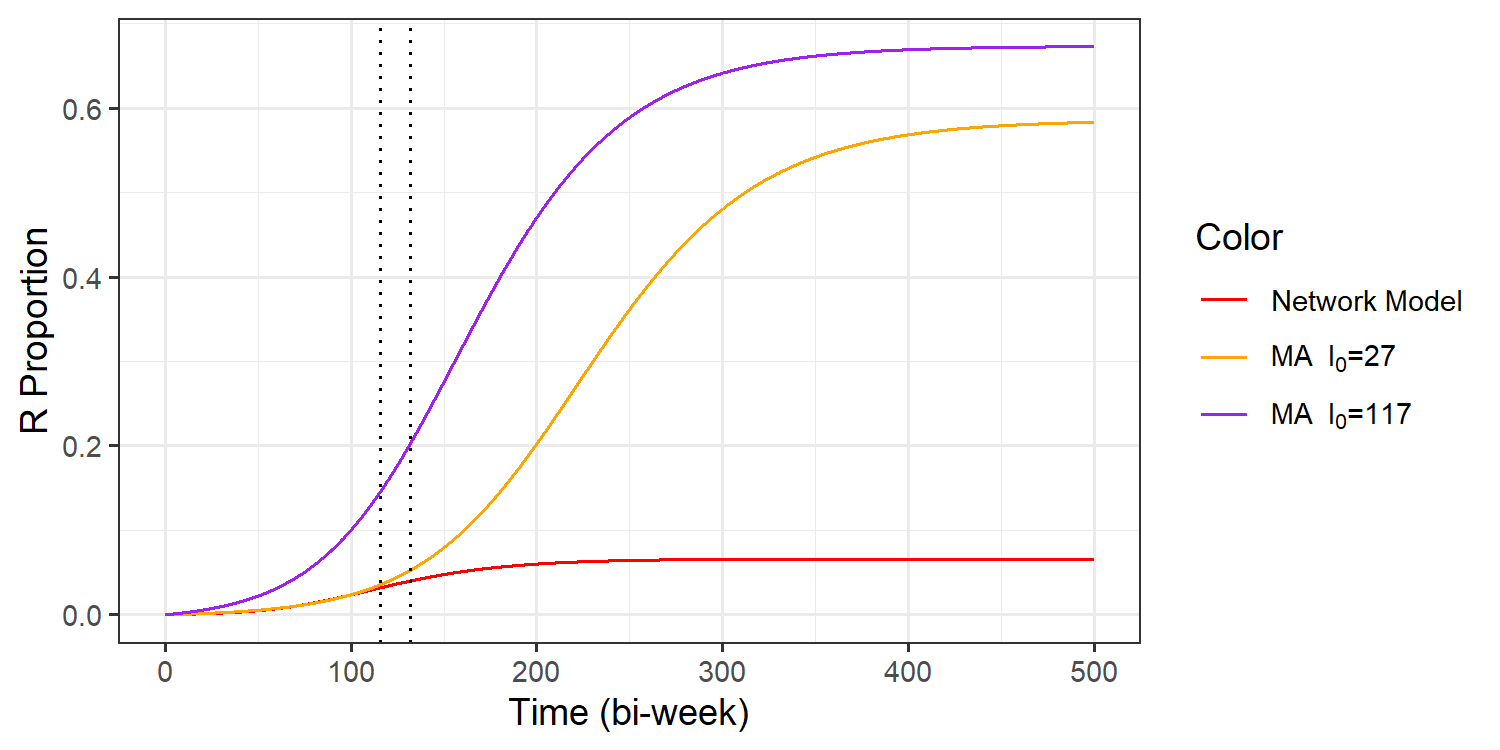}
         \caption[MASIR R Tendency]{Model trajectories for the recovered compartment $R(t)$}
    \end{subfigure}
    \begin{subfigure}[b]{0.49\textwidth}
         \centering
         \includegraphics[width=1\textwidth]{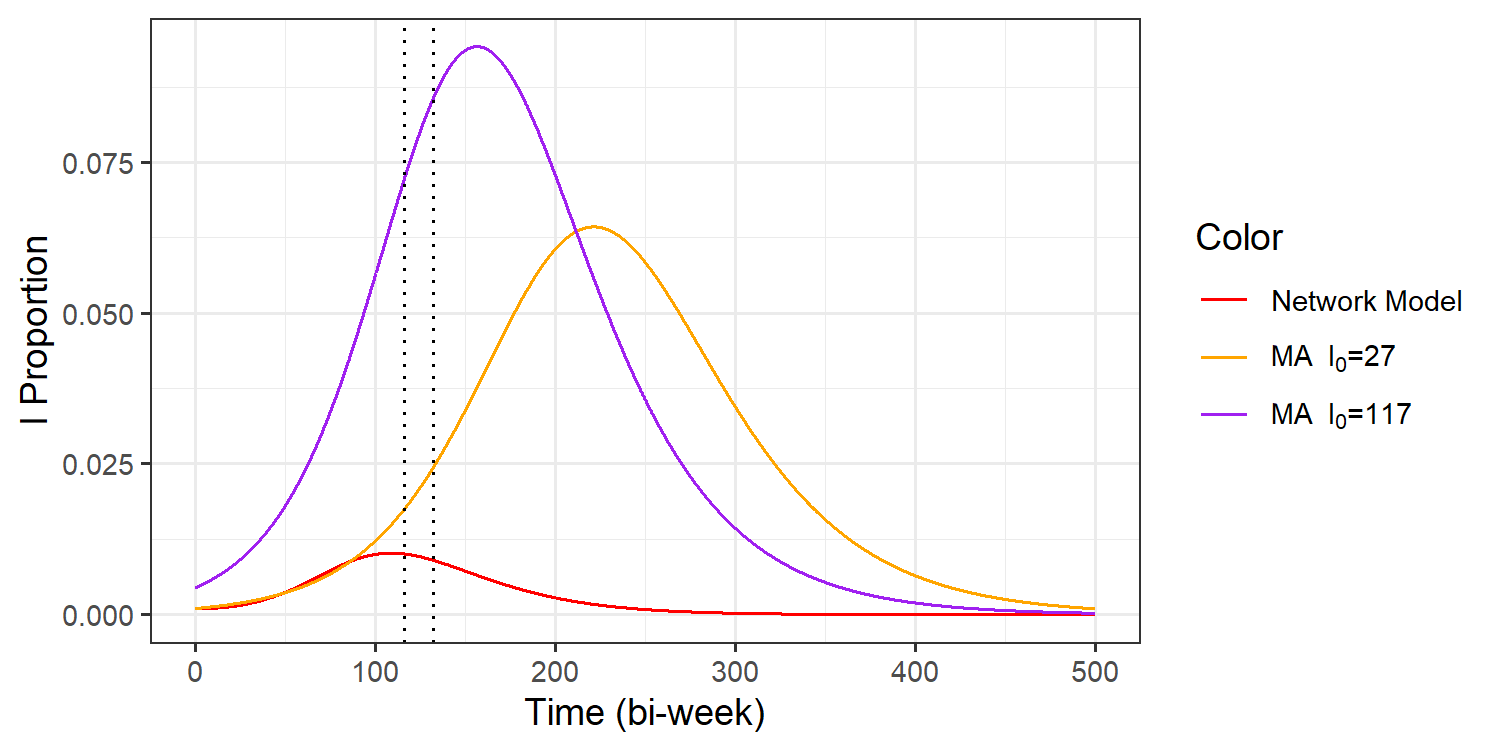}
         \caption[MASIR I Tendency]{Model trajectories for the infectious compartment $I(t)$}
    \end{subfigure}
    \caption{The prediction trajectories for $R(t)$ and $I(t)$ until the end of outbreak. The trajectories of the two MA-SIR models and network-SIR model are shown. For each sub-figure, the first dotted vertical line is when the POCT is implemented and the second is the end of time in current collected data.}
    \label{fig: MASIR Tendency}
\end{figure}

We see from Table~\ref{table: MASIR Init117} that in both cases the MA-SIR model tends to predict a larger final infected size that is more than nine times greater than what we get from the network-SIR model. 
These MA-SIR models actually predict that more than half of the target population will be infected.
We can further see such behaviour from the comparison of the dynamics in Figure~\ref{fig: MASIR R Cumulative} and Figure~\ref{fig: MASIR Tendency}.
It is clear that the homogeneous mixing MA assumption has very strong consequences.

We also note that in Figure~\ref{fig: MASIR Init117} (a) and Figure~\ref{fig: MASIR Init27} (a), the boundary of the grey area on right side of the heat maps is no longer a relatively smooth curve like in Figure~\ref{fig: MSV Init27} for network models.
As mentioned in Section~\ref{sec: result}, the grey areas on the left for all three cases reflect prediction trajectories where $X(t)$ is lower than $\hat{n}(t)-1$ contributing to a likelihood of zero.
This is also true for the right grey area for network case, however since the MA-SIR models tends to estimate more cases and more rapid biweekly incidence while some of the case counts from data are low, the corresponding PDF $\hat{f}_{CB}$ can go very close to zero.
This becomes more of an issue as $\beta$ increases (going to the right on the heat maps).
Numerical underflow eventually causes $\hat{f}_{CB}=0$ and $\ell = -\infty$, which is also plotted in grey.
Our investigation of this issue says that the parts of this grey area on the right side of the heat maps for MA-SIR models is still feasible but with extremely low probability or likelihood.
Moreover, with further empirical investigation, the irregular boundary is combined by several relatively smooth curves, each corresponding to issues with a low value data point.

\begin{figure}[h]
    \begin{center}
        \includegraphics[width=0.95\textwidth]{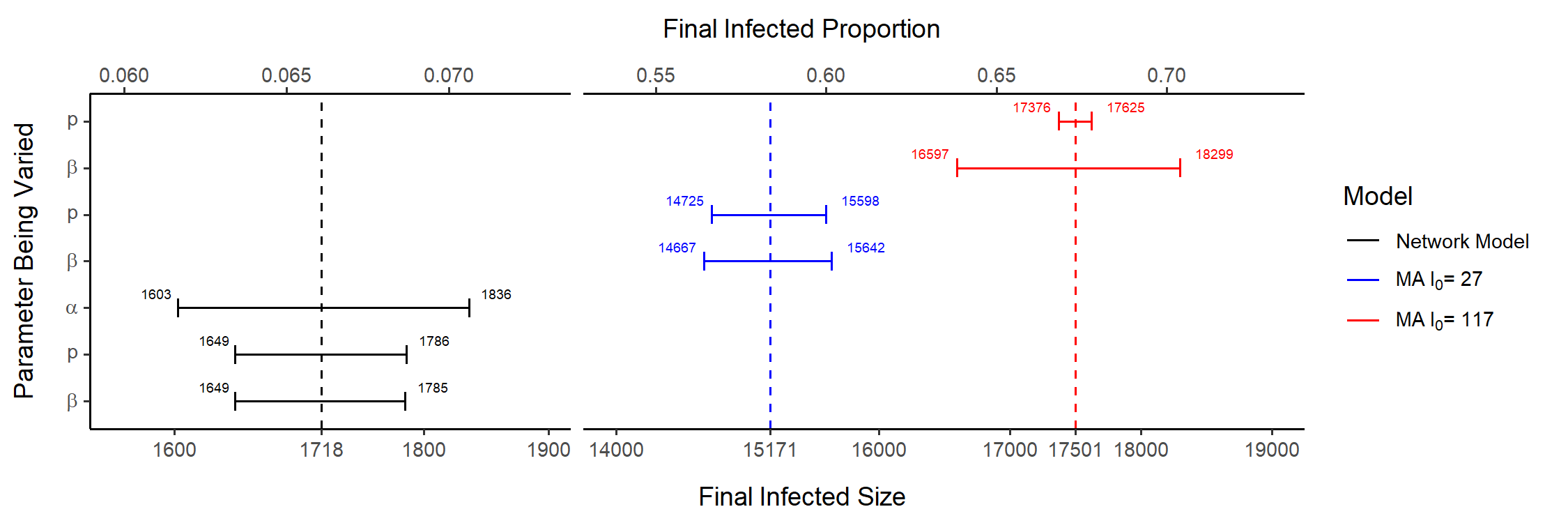}
    \end{center}
    \caption[Sensitive Analysis]{Sensitivity analysis of final infected size $R(\infty) \times N$ and corresponding proportion $R(\infty)$ for the network-SIR and MA-SIR models. 
    The range of values used when varying $\beta$ and $p$ is their confidence intervals in Table~\ref{table: MSV Init27} and \ref{table: MASIR Init117}. The range used for $\alpha$ is given by $\hat{\alpha} \times (100\% \pm 1\%)$.
    The dashed vertical line represent the final infected size from optimal fitted model.}
    \label{fig: Sensitivity}
\end{figure}

We present the results of a sensitivity analysis of the final infected size $R(\infty) \times N$ with respect to some of the fitted parameters in Figure~\ref{fig: Sensitivity}.
For the network model, we consider the impact of the per-infected transmission rate $\beta$, reporting probability $p$ and the power law parameter of degree distribution $\alpha$.
For the MA-SIR models, we show the impact of $\beta$ and $p$. 
We perturb one parameter each time while keep all other parameters fixed and calculate the resulting final infected size accordingly.
Note that $p$ and $\gamma$ are still tied by the relationship as described in Section~\ref{sec: reporting}.
The range of values used for $\beta$ and $p$ is given by their confidence intervals given in Table~\ref{table: MSV Init27} and \ref{table: MASIR Init117}.
The range used for $\alpha$ is given by $\hat{\alpha} \times (100\% \pm 1\%)$.
We did not include sensitivity to $I_0$ here since this can already be observed in Figures~\ref{fig: MSV I0} and \ref{fig: MASIR I0}.

\section{Point of Care Test and Treat}
\label{sec: POCT}

Typically, syphilis diagnosis requires a suspected case to identify symptoms or be informed from a known case of a possible infection, find an appropriate clinic, access health services with a valid health insurance number, attend the initial appointment for serological testing and come back to the clinic days later to learn the results of the test. 
Each step in this process makes it less  accessible for underserved community members to be diagnosed and treated, and more likely to be lost to follow up during the early stages when the infection is most infectious.

As a response to the reported local outbreak, KFL\&A PH implemented a rapid test and treatment protocol using a rapid syphilis antibody POCT. 
The medical directive was to treat suspected cases based on POCT results immediately, significantly reducing the TAT time and allowing testing without clinic or laboratory environments.
This protocol is employed by outreach nurses, and implemented at various sites, including community-based organizations and advertised outreach events (Blitzes), making it more accessible to the population at greatest risk.
Since this POCT was only recently approved by Health Canada, it is still in the very early stages of real-world investigation~\cite{MackrellEtAl:2024}.
Using our existing model, we aimed to support this effort by building prediction models with some conjectures to illustrate the potential impact of POCT.

We assume that the test is accurate and would be applied to every case reported.
The network-SIR model in Section~\ref{sec: result} is considered our baseline model, for the scenario when POCT is not implemented.
We consider different scenarios where we increase coverage of POCT implementation, leading to an increase in the reporting probability while also decreasing the test TAT for all reporting cases.
We consider the following scenarios:
\begin{enumerate}
    \item Reporting probability is increased to $\Tilde{p}_1=\hat{p}+0.05$ and infectious duration of reported cases is decreased by nine days.
    \item Reporting probability is increased to $\Tilde{p}_2=\hat{p}+0.10$ and infectious duration of reported cases is decreased by nine days. 
    \item Reporting probability is increased to $\Tilde{p}_1=\hat{p}+0.15$ and infectious duration of reported cases is decreased by nine days.
\end{enumerate}

The decrease in the infectious duration is done by reducing by $\frac{9}{14}$ bi-week units the data point for the percentiles of primary infections (from four bi-weeks to $\frac{47}{14}$ bi-weeks) and secondary stage (from ten bi-weeks to $\frac{131}{14}$ bi-weeks when fitting the relationship between the reporting probability $\Tilde{p}_i$ and the corresponding modified recovery rate $\Tilde{\gamma}_i$ for scenarios $i=1,2,3$ using the procedure discussed in Section~\ref{sec: reporting}.
The per-infected transmission rate $\beta$ is left unchanged.

\begin{figure}[htbp]
    \centering
    \begin{subfigure}[b]{1\textwidth}
         \centering
         \includegraphics[width=1\textwidth]{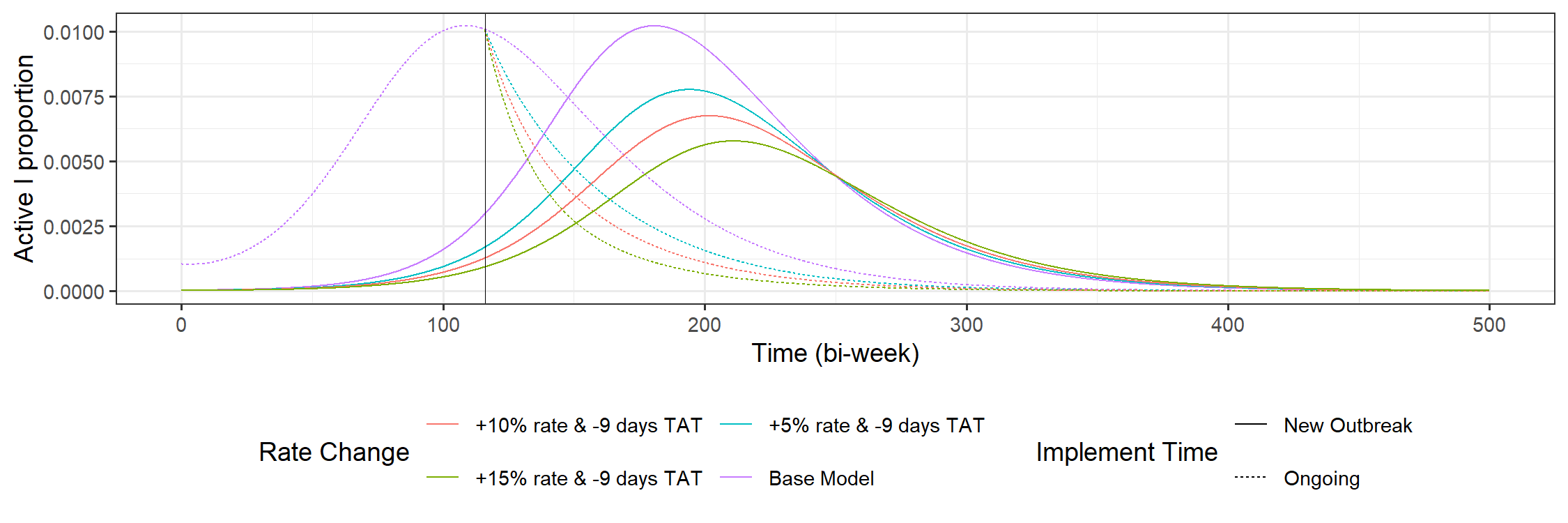}
         \caption[POCT I Tendency]{The prediction trajectories for $I(t)$.}
    \end{subfigure}
    \begin{subfigure}[b]{1\textwidth}
         \centering
         \includegraphics[width=1\textwidth]{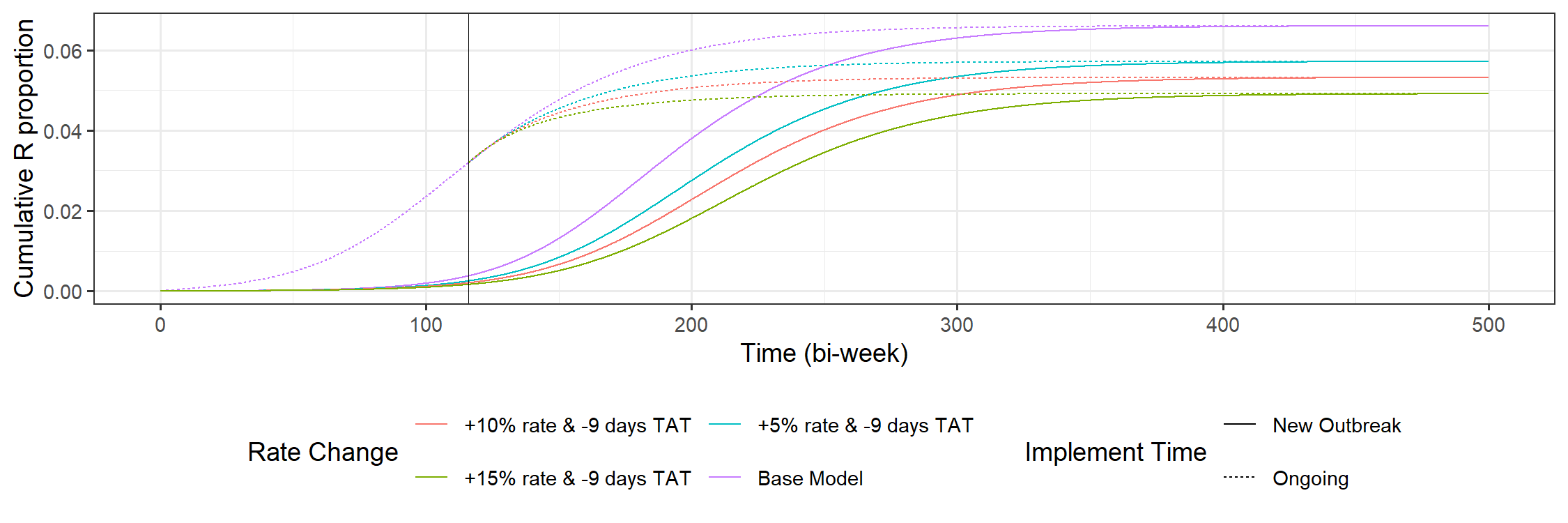}
         \caption[POCT R Tendency]{The prediction trajectories for $R(t)$.}
    \end{subfigure}
    \caption{Comparison of the trajectories under three different scenarios and two different settings of POCT implementation. The trajectories of the baseline network-SIR model is also included for comparison.
    The vertical lines in both sub-figures denote when POCT was actually implemented by KFL\&A Public Health.}
    \label{fig: POCT}
\end{figure}

In Figure~\ref{fig: POCT} we show the effect of scenarios 1--3 in two different settings. First, we consider the effect of POCT on an entirely new epidemic by generating trajectories as in Figure~\ref{fig: MASIR Tendency} using the same transmission rate $\beta$ as the baseline model but with $I_0=1$, new recovery rate $\gamma=\Tilde{\gamma}_i$ and new reporting probability $p=\Tilde{p}_i$ for each of the scenarios.
The resulting trajectories are shown in in Figure~\ref{fig: POCT} as solid curves.

The second setting we consider is the effect of POCT on the ongoing outbreak in KFL\&A by assuming there was a switch in the recovery rate and reporting probability on June 2023 ($t=116$, after the end of the fitted data). 
To do this we stopped the baseline model at $t=116$ and then restarted to generate trajectories for the next time period with the new reporting probability $p=\Tilde{p}_i$ and new recovery rate $\gamma=\Tilde{\gamma}_i$ for each scenario.
The results are shown in Figure~\ref{fig: POCT} as dotted curves.


From the theory discussed in literature~\cite{MillerSlimVolz:2012,ZhaoMagpantay:2024}, we know that the final size of the epidemic is determined by the equilibrium of the system \eqref{eqn: CM ODE}, and thus independent of the setting.
We see this in Figure~\ref{fig: POCT}. The values of the equilibrium under different POCT implementation is presented in Table~\ref{table: POCT}.

\vspace{-5mm}
\begin{center}
\begin{table}[htbp]
\centering    \caption{Model assumptions and predicted final epidemic sizes for the different POCT implementation scenarios. Rates ($\Tilde{\gamma}$ and $\beta$) have unit biweek$^{-1}$. }
    \label{table: POCT}
        \begin{tabular}{|C{0.6in}|C{2.0in}|C{0.7in}|C{0.7in}|C{0.7in}|C{0.7in}|}
            \hline
            Notation & Definition & Baseline & POCT 1 & POCT 2 & POCT 3
             \\
            \hline
            $N$ & Population/Network Size & \multicolumn{4}{c|}{26,000} 
            \\
            \hline
            $\alpha$ & Power law parameter of degree distribution & \multicolumn{4}{c|}{1.738004}
            \\
            \hline
            $I_0$ & Initial active case count &  \multicolumn{4}{c|}{1}
            \\
            \hline
            $\beta$ & Uniform per-infected transmission rate & \multicolumn{4}{c|}{0.01926}
            \\
            \hline
            $\Tilde{p}$ & Uniform reporting probability & 0.2755 & 0.3255 & 0.3755 & 0.4255
            \\
            \hline
            $\Tilde{\gamma}$ & Optimal uniform per-infected recovery rate & 0.05192518 & 0.05652490 & 0.05874144 & 0.06111252
            \\
            \hline
            $R_{0}$ & Basic reproductive number & 1.8548 & 1.7088 & 1.6464 & 1.5845
            \\
            \hline
            $R(\infty)$ & Final infected proportion & 0.06606 & 0.05722 & 0.05325 & 0.04917
            \\
            \hline
            $R(\infty) \times N$ & Final infected size & 1717.588 & 1487.721 & 1384.389 & 1278.466
            \\
            \hline
            N/A & Final infection size relative to baseline model & 100.00\% & 86.62\% & 80.60\% & 74.43\%
            \\
            \hline
            
        \end{tabular}
    \end{table}
\end{center}
\vspace{-5mm}

We see from Table~\ref{table: POCT} that increasing reporting probability by $0.05$, together with the TAT reduction on reported cases has a significant impact on the final infected size, with a reduction of $13.38\%$ compared to the baseline model.
After the first $5\%$ increase in reporting probability, the other two additional increases in reporting probability each lead to an additional decrease of approximately 6\% of the final infected size.
Since the final infected sizes are independent of initial conditions, these results suggest that there can be a great advantage in implementing POCT even if the intervention was started after the peak of the infection. 
The POCT can still efficiently decrease the final infection size even if it only adds a $5\%$ increase in the reporting probability.

To illustrate the benefit and efficiency of the rapid test and treat strategy, we present a simple economic analysis here based on the POCT 1 scenario. 
Our model prediction in Table~\ref{table: POCT} suggests that increasing the reporting probability by 5\% and using the rapid test and treat strategy in the KFL\&A area would lead to an average of $1487.721$ infections (0.05722 of the target population), which is about $230$ less infections than the baseline. 
Since this scenario requires an overall $0.3255$ reporting probability, an expected total of $0.3255\times 1487.721 \approx 484.3$ of cases would need to be reported. The infected fraction of the population under this scenario is 0.05722 so we assume that strategies can be devised to find this many cases after testing approximately $484.3 \div 0.05722 \approx 8463$ individuals.
Based on KFL\&A PH estimates, the average cost of each POCT is \$45 CAD (\$10-\$16 for test kit and \$25-\$40 for average nursing time and associated costs) so the POCT1 scenario has an estimated cost of $8643
\times 45 = 388,935$ CAD.
On the other had, Chesson and Peterman~\cite{ChessonPeterman:2021} estimated that the discounted lifetime cost per infection of syphilis was $1,190$ USD in the United States in 2019.
After adjusting for the the exchange rate and inflation, we projected an average discounted lifetime cost of $2,000$ CAD for KFL\&A PH in 2024.
Even without accounting for the cost of the baseline scenario using traditional testing, the decrease of just 230 infections from baseline in the POCT1 scenario would already yield a savings of $460,000$ CAD, enough to cover the entire estimated cost of the POCT1 strategy. 
This very conservative economic analysis favours implementing POCT1, assuming the test is accurate.
The estimated economic benefits could be even higher if we also consider (1) the public health resources that are saved when attempting to track ``lost'' cases and (2) the reduction in average infection time, since earlier treatment of syphilis would lead to better health outcomes~\cite{PHACSyphilis}.

From Figure~\ref{fig: POCT} (a), another potential effect of POCT would be ``flattening the curve'' and delaying the peak of the outbreak, which creates more time for public health professionals to implement more time-consuming interventions such as awareness campaigns to promote safe sex to suppress $\beta$.

\section{Conclusions}

In this paper we applied percolation-based network epidemiology models to model the transmission of syphilis in the KFL\&A PH area.
In Section~\ref{sec: model}, we described the construction of the baseline network-SIR model from the collected data, including assumptions we made on the size of the target population and the role of risk factors in scaling up the reported number of contacts from the data.
We discussed how we fit a power law distribution to the scaled sample degree distribution and presented the equations used for the dynamics of the model.
In Section~\ref{sec: reporting} we discussed the connection between reporting probability and recovery rate, tying them together by fitting the CDF of an exponential distribution for the infectious period.

In Section~\ref{sec: MLE} we presented the parameters of the network-SIR model are determined  using maximum likelihood. We also presented the expected trajectory and final epidemic size.
In Section~\ref{sec: MA compare} we compared the results of the network-SIR model with those of traditional MA-SIR models. 
The comparison clearly illustrates the advantages of percolation-based network models and the limitations of MA-SIR models when modeling disease outbreaks wherein network structure is very important. 
The MA-SIR model tends to estimate far larger final epidemic sizes than the network-SIR model. 

Finally in Section~\ref{sec: POCT}, we modeled the impact of rapid test and treat intervention under different scenarios. For each scenario we also considered two different settings, reflecting different timings of the implementation of POCT.
The results shows that the POCT implementation can be very impactful in suppressing the final outbreak size, even with relatively low level of reporting.
Moreover, the model indicates that if the POCT is implemented before the outbreak, it will flatten the curve and delay the peak of infection, allowing  public health workers time to implement other control strategies.
We included a conservative economic analysis of the POCT strategy which supports the use of POCT.

As discussed in Section~\ref{sec: model}, our model employs many assumptions due to limitations in data and resources.
The estimates and predictions we generated should only be used to compare the scenarios considered and the models used (MA-SIR versus network-SIR). 
We believe this exercise is still useful as a tool for supporting public health in modeling the spread of syphilis and the effect of different control efforts.




We plan to conduct further investigation on this work from multiple perspectives.
One goal is to remove some of our simplifying assumptions.
For example, we would be interested in modeling reinfection by using a susceptible-infectious-recovered-susceptible (SIRS) model and removing (H5).
Another aim for future work concerns the construction of the network. We would be interested in comparing our scaled degree distribution to real-life degree distributions if more data were available through behaviour surveys in the underserved communities.
This can improve our estimates of the size of the target population as well as the network degree distribution, and the results would also apply to modeling other types of sexually transmitted infections.
In addition, for simplicity we had assumed that the sensitivity of the POCT strategy is perfect, but empirical evidence from the current intervention process indicates otherwise.
We plan to extend our economic analysis to explore imperfect sensitivity in future work.

The rapid test and treat intervention has been expanded to include eight public health units across Ontario, known as the SPRITE Study~\cite{MackrellEtAl:2024}
Implementation data will be collected using the Reach, Effectiveness, Adoption, Implementation, and Maintenance (RE-AIM) Framework, following a community-based participatory approach.
The data will be used to refine our estimates of the impact of rapid test and treat better, without relying on hypothetical scenarios discussed in Section~\ref{sec: POCT}.
We also aim to expanding our framework to other communities with different social structures or to other diseases with different transmission dynamics.


\section*{Acknowledgements}
We thank Duy A. Dinh for his help with the literature review. 

\section*{Funding Statement}
This work was supported by the Canadian Institutes of Health Research (CIHR) through a Catalyst (SR8 190795), Operating (AS1-192619), and Knowledge Mobilization (EKS 193138) granted to the NPA SS. 
FMGM also acknowledges the support of the Natural Sciences and Engineering Research Council of Canada (NSERC), Mathematics for Public Health and the Canadian Foundation for Innovation.

\newpage
\bibliographystyle{plain}
\bibliography{Bibliography.bib}

\newpage
\include{supplement}

\end{document}

%% file: supplement.tex
\setcounter{equation}{0}
\setcounter{figure}{0}
\setcounter{table}{0}
\setcounter{section}{0}
\setcounter{page}{1}
\makeatletter
\renewcommand{\theequation}{S\arabic{equation}}
\renewcommand{\thefigure}{S\arabic{figure}}
\renewcommand{\thetable}{S\arabic{table}}
\renewcommand{\thesection}{S\arabic{section}}

\begin{center}
{\huge \sc   Supplementary Information}
\end{center}

\section{Adjustment of number of contacts based on age and risk factors}
\label{sec:rescaling_contacs}

To do this,
we assumed that age and aggregate risk factors (aggregation discussed later and presented in Table~\ref{table: age factors}) are independent.
Moreover, we assume that the adjustments corresponding to each aggregate risk factor group are independent and additive.
Therefore, for an individual of age $x$ that reported $n$ contacts, the adjusted degree $\hat{n}$ is given by,
\begin{equation}
\label{eqn: scale}
\hat{n} = \Big\lfloor n \times (1+\mathbb{F}_{age} (x)) \times (1 + \sum_{i=1}^{10} \mathbb{1}_{risk_i} \times r_i )\Big\rfloor.
\end{equation}
Here $\lfloor \cdot \rfloor$ represents the floor function, so that for any real number $y$,
\begin{equation}
    \lfloor y \rfloor = \max \{k \in \mathbb{Z}|k \leq y\}.
\end{equation}
The expression $\mathbb{1}_{risk_i}$ is an indicator function that equals $1$ if the case has a risk factor that belongs to risk group $i$, and is zero otherwise.
We note that we only consider risk groups here. 
If multiple risk factors within one risk group appears for an individual then we use the correction for that risk group only once.

Finally, the function $\mathbb{F}_{age} (x)$ is a piece-wise function depending on age defined as,
\begin{equation}
    \mathbb{F}_{age} (x) = \begin{cases}
        a_1, \text{ if }x \in [18,20), \\
        a_2, \text{ if }x \in [20,30), \\
        a_3, \text{ if }x \in [30,40) ,\\
        a_4, \text{ if }x \in [40,50) ,\\
        a_5, \text{ if }x \in [50,\infty). \\ 
    \end{cases}
\end{equation}
The scaling parameters $a_1,\dotsc,a_5$ associated with age are found in Table~\ref{table: age factors} and the scaling parameters $r_1,\dotsc,r_{10}$ associated with risk factors are found in Table~\ref{table: risk factors}. 
These were established after consultation with public health professionals, front-line public heath nurse, infectious disease physicians, public health epidemiologist and other collaborators/authors, based on their observations, professional experience and knowledge of the literature on syphilis~\cite{AhoEtal:2022,PHOCongenital:2022}.
Estimates of these parameters were considered specifically for the KFL\&A target population and we present some discussion of how they were arrived at below.
For this work we only aim to provide a practical framework based on reported cases data, professional observation and opinions, which are accessible to knowledge-users of similar projects.
The procedure can be easily adapted to accommodate changes in the estimates of risk and age factors. 
We conducted sensitivity analysis using different scaling values and the results do not appear to be highly sensitive to the exact values of these factors.

\begin{center}
    \begin{table}[htbp]
    \centering
\caption{Age groups of adult cases with reported numbers of contacts. Values of scaling parameters were determined by consultation with public health professionals.}
    \label{table: age factors}
        \begin{tabular}{|C{1.0in}|C{1.0in}|C{0.8in}|C{0.8in}|C{0.8in}|}
            \hline
            \multicolumn{3}{|c|}{Age Groups} & \multicolumn{2}{c|}{Scaling Parameters}
            \\
            \hline
            Age range & Risk Level & Proportion & Symbol & Value
            \\
            \hline
            $[18,20)$ & Medium & 1.25\% & $a_1$ & 0.150
            \\
            \hline
            $[20,30)$ & High & 26.97\% & $a_2$ & 0.250
            \\
            \hline
            $[30,40)$ & Medium & 40.66\% & $a_3$ & 0.150
            \\
            \hline
            $[40,50)$ & Low-Medium & 21.16\% & $a_4$ & 0.050
            \\
            \hline
            $[50,\infty)$ & Low & 9.96\% & $a_5$ & 0.025
            \\
            \hline
        \end{tabular}
    \end{table}
\end{center}

\begin{center}
    \begin{table}[htbp]
    \centering
\caption{Risk factors and aggregated risk factor groups of  adult cases. Scaling parameter values are determined by consultation with public health professionals.}
\label{table: risk factors}
\begin{tabular}{|C{2.4in}|C{0.4in}|C{0.9in}|C{0.7in}|C{0.7in}|}
            \hline
            \multicolumn{3}{|c|}{Risk Factor Groups} & \multicolumn{2}{c|}{Scaling Parameters}
            \\
            \hline
            Factor Description & Count & Group & Symbol & Value 
            \\
            \hline
            Anonymous Sex & 42 & \multirow{3}{0.7in}{\centering Anonymous Sex} & \multirow{3}{*}{$r_1$} & \multirow{3}{*}{$0.6$} 
            \\
            \cline{1-2}
            Met contact through internet & 5 & & &
            \\
            \cline{1-2}
            Bath house & 1 &  &  &
            \\
            \hline
            Repeat STI & 55 & \multirow{5}{0.7in}{\centering STI} & \multirow{5}{*}{$r_2$} & \multirow{5}{*}{$0.2$} 
            \\
            \cline{1-2}
            Co-Infection with other STI & 10 & & &
            \\
            \cline{1-2}
            HIV Status & 5 &  &  &
            \\
            \cline{1-2}
            PrEP for HIV & 4 &  &  &
            \\
            \cline{1-2}
            HIV positive contact & 1 &  &  &
            \\
            \hline
            Pregnant & 16 & \multirow{8}{0.7in}{\centering Pregnancy Related} & \multirow{8}{*}{$r_3$} & \multirow{8}{*}{$0.0$} 
            \\
            \cline{1-2}
            Tested during first Trimester & 4 & & &
            \\
            \cline{1-2}
            Treated $>4$ weeks prior to delivery  & 2 &  &  &
            \\
            \cline{1-2}
            Received prenatal care & 2 &  &  &
            \\
            \cline{1-2}
            Tested at delivery & 1 &  &  &
            \\
            \cline{1-2}
            Tested at 28-32 Weeks & 1 &  &  &  
            \\
            \cline{1-2}
            Tested $>4$ weeks prior to delivery & 1 &  &  &  
            \\
            \hline
            Correctional facility & 20 & \multirow{1}{0.7in}{\centering Prison} & \multirow{1}{*}{$r_4$} & \multirow{1}{*}{$0.2$} 
            \\
            \hline
            Under-housed/Homeless & 28 & \multirow{1}{0.7in}{\centering SES} & \multirow{1}{*}{$r_5$} & \multirow{1}{*}{$0.0$} 
            \\
            \hline
            No condom used & 55 & \multirow{4}{0.7in}{\centering Sex Behavior} & \multirow{4}{*}{$r_6$} & \multirow{4}{*}{$0.0$}
            \\
            \cline{1-2}
            More than one sex contacts & 22 &  &  & 
            \\
            \cline{1-2}
            Shared sex toys & 1 &  &  & 
            \\
            \cline{1-2}
            New sex contact in past 2 months & 23 &  &  & 
            \\
            \hline
            Sex trade worker & 11 & \multirow{2}{0.7in}{\centering Sex Trade} & \multirow{2}{*}{$r_7$} & \multirow{2}{*}{$1.0$} 
            \\
            \cline{1-2}
            Sex with sex trade worker & 6 &  &  & 
            \\
            \hline
            Sex with same sex & 45 & \multirow{1}{0.7in}{\centering Same Sex} & \multirow{1}{*}{$r_8$} & \multirow{1}{*}{$0.6$} 
            \\
            \hline
            Injection drug use & 38 & \multirow{6}{0.7in}{\centering Substance} & \multirow{6}{*}{$r_9$} & \multirow{6}{*}{$0.3$} 
            \\
            \cline{1-2}
            Inhalation drug use & 43 &  &  & 
            \\
            \cline{1-2}
            Judgement impaired by alcohol/drugs & 8 &  &  & 
            \\
            \cline{1-2}
            Shared other drug equipment & 6 &  &  & 
            \\
            \cline{1-2}
            Sex for drugs/shelter/food & 3 &  &  & 
            \\
            \cline{1-2}
            Shared needles & 2 &  &  & 
            \\
            \hline
            Tattoo and piercing & 5 & \multirow{5}{0.7in}{\centering Unknown} & \multirow{5}{*}{$r_{10}$} & \multirow{5}{*}{$0.0$} 
            \\
            \cline{1-2}
            Travel outside province & 4 &  &  &
            \\
            \cline{1-2}
            Other social venue & 1 &  &  & 
            \\
            \cline{1-2}
            Unknown & 248 &  &  & 
            \\
            \cline{1-2}
            Other & 7 &  &  & 
            \\
            \hline
        \end{tabular}
    \end{table}
\end{center}

Applying equation~\eqref{eqn: scale} to all data with number of contacts yielded an adjusted integer-valued degree sequence.
This is an estimate of more realistic numbers of sexual contacts of the target population for the time period we are interested in.
A comparison of the original degree sequence and the adjusted degree sequence is shown in Figure~\ref{fig: Data Adj}.